\documentstyle[aps,twocolumn,psfig,verbatim,citesort,amstex]{revtex}

\begin{document}

\draft

\bibliographystyle{prsty}
\title{Quantum-classical correspondence in the hydrogen atom in weak external fields}

\author{Paolo Bellomo$^{1}$, C. R. Stroud, Jr.$^{1}$, 
David Farrelly$^2$ and  T. Uzer$^3$}

\address{$^1$Rochester Theory Center for Optical Science and Engineering and 
The Institute of Optics, University of Rochester,
Rochester, NY 14627-0186, USA. \\
$^2$Department of Chemistry and Biochemistry, Utah State University, 
Logan, Utah 84332-0300, USA. \\
$^3$School of Physics, Georgia Institute of Technology, 
Atlanta, Georgia 30332-0430, USA.}

\date{\today}

\maketitle

\begin{abstract}
The complex processes leading to the collisional population of
ultralong-lived Rydberg states with very high angular momentum can be
explained surprisingly well using classical mechanics. In this article, we
explain the reason behind this striking agreement between classical theory
and experiment by showing that the classical and quantum dynamics of
Rydberg electrons in weak, slowly varying external fields agree beyond the
mandates of Ehrenfest's Theorem. In particular, we show that the
expectation values of angular momentum and Runge-Lenz vectors in hydrogenic
eigenstates obey exactly the same perturbative equations of motion as the
time averages of the corresponding classical variables. By time averaging
the quantum dynamics over a Kepler period, we extend this special
quantum-classical equivalence to Rydberg 
wave packets relatively well localized in energy.
Finally, the perturbative equations hold well also for external fields beyond the 
Inglis-Teller limit, and in the case of elliptic states,
which yield the appropriate quasiclassical initial conditions,
the matching with classical mechanics is complete.
\end{abstract}


\section{Introduction}

In the past few years 
new experimental techniques have made possible the study of the dynamics of
atoms or molecules in which an electron is promoted to a very high energy 
state, where it is only weakly bound to the core {\cite{amato_96}}. 
These high energy states can be described by 
approximately hydrogenic wave functions with 
very large principal quantum 
numbers ($n \gtrsim 100$) {\cite{gallagher_94,herzberg_87}}.
The atoms (or molecules) in which a valence electron 
is promoted to such high-{\em n} 
states are generically called ``Rydberg" atoms,
because the energy levels of the excited electron are well described by a 
Rydberg-like formula {\cite{gallagher_94}}, and their highly
energetic electron is known as a Rydberg electron.
In such systems the weakly bound Rydberg electron 
resides mostly at 
an immense distance from the atomic core, a distance so large that if
Rydberg atoms were solid, they would be just about visible to the naked eye. 
Laboratory-scale external fields, and even weak stray electric 
fields {\cite{gross_86a,vrakking_95a,vrakking_95b,vrakking_95c,horsdal_98a}}, 
become then comparable to the atomic
Coulomb field sensed by the Rydberg electron, 
so that the dynamics of the electron can be probed with accuracy, and also
fundamental dynamical properties such as
quantum manifestations of chaos 
{\cite{eck,hasegawa_89,gutzwiller_90,nakamura_93,haake_91,koch_95},
can be studied experimentally.

To a very good approximation,
the dynamics of Rydberg electrons is hydrogenic.
More precisely, small deviations from the purely hydrogenic 
eigenenergies are introduced by the interaction between the far flung electron 
and the electronic cloud around the atomic or molecular core.
These deviations are quantified by the quantum defect $\delta_{\ell}$, 
which enters in the formula for the energy levels as a correction to 
the principal quantum number $n$ {\cite{gallagher_94,freeman_76}}.
However, the quantum defect becomes rapidly negligible as the angular momentum 
of the electron increases.
In fact, more complex atoms are often used as experimental substitutes for
hydrogen, since it is much easier to excite their valence electron
to a Rydberg state, and yet the field sensed by the Rydberg electron does
not differ much from a pure Coulomb field. 
Therefore many recent investigations 
of Rydberg electrons in alkali atoms
have really probed the dynamics of quasiclassical
electrons in an essentially coulombic potential.
These accurate experimental results
have led to a renewed theoretical interest in the hydrogen atom in external
fields in the limit of large quantum numbers,
{\cite{hasegawa_89,demkov_70,demkov_74,turgay_94a,%
turgay_94b}}
which has become one of the paradigmatic models for 
the study of quantum chaos
{\cite{eck,hasegawa_89,gutzwiller_90,nakamura_93,haake_91,koch_95},
and of quantum-classical correspondence in general.

Since the degeneracy of a hydrogenic $n$-manifold grows as $n^2$, a fully quantum 
treatment of the dynamics of Rydberg electrons
poses formidable challenges even to the most advanced computers. Therefore
classical mechanics is often the only practical way to 
study such systems, under the assumption that for large $n$'s classical
and quantum predictions should somehow converge. 
Interestingly, however, recent experimental and theoretical work on electronic
wave packets in hydrogenic systems
{\cite{%
carlos_86a,carlos_91a,carlos_94a,carlos_94b,carlos_94c,%
carlos_95a,carlos_96a,carlos_97a,%
perelman_89,%
nauenberg_89a,nauenberg_90a,nauenberg_94a,%
eberly_94a,eberly_95a,eberly_96a,eberly_96b,%
turgay_95a,turgay_96a,turgay_97a,david_97a,turgay_98a,%
bialynicki_97a,delande_95a,delande_95b,klauder_96a,myself_11}
have shown very clearly that the
quantum mechanical properties of the Rydberg electron are 
essential to the dynamics of the wave packet,
even in the large-{\em n} regime. For example, 
the observation of fractional revivals
can be explained only by the quantized spectrum  of the Hamiltonian
{\cite{carlos_86a,carlos_91a,carlos_94a,carlos_95a,carlos_97a,perelman_89,nauenberg_89a}}
in spite of the very large principal quantum numbers involved, which seems to 
suggest that large quantum numbers are not sufficient to ensure 
the accuracy of a purely classical description of Rydberg dynamics.

On the other hand, classical mechanics yields
surprisingly accurate results for
the problem of the hydrogen atom in weak, slowly 
varying external fields; i.e., when the classical
electron still moves, to a good approximation, 
along a Kepler ellipse, and the semimajor axis of the ellipse
(or, equivalently, the Kepler energy of the electron) remains unchanged.
In particular, recent classical, perturbative calculations 
{\cite{gross_86a,nauenberg_93a,myself_7,myself_9,myself_8,myself_6,carlos_98a}} 
have succeeded in explaining several diverse experimental
results with astonishing accuracy, ranging, for example,
from the ultralong lifetimes of the molecular 
high-$n$ states employed in
zero-electron-kinetic-energy (ZEKE) spectroscopy
{\cite{reiser_88,zeke_0,zeke_1,zeke_5,chupka_93a,chupka_93b,%
vrakking_95a,vrakking_95b,vrakking_95c,myself_7,myself_9,merkt_97a}},
to the intrashell transitions 
induced in alkali Rydberg atoms by slow ion-Rydberg collisions
{\cite{macadam_81,macadam_85,macadam_88a,macadam_93,macadam_95,myself_7,myself_9}}.
The same classical approach also explains
the anomalous autoionization lifetimes of Rydberg electrons 
in circularly polarized microwave fields
{\cite{gallagher_97,myself_8}}
and the dynamics and stability of circular Rydberg states (i.e., states with
maximum angular momentum) in weak, slowly rotating electric fields
{\cite{gross_86a,myself_6}}.
Clearly, the success of the classical treatment must stem from some special equivalence
between classical and quantum predictions in the hydrogen atom in
weak external fields: in fact, such equivalence
has been already simply observed (but not explained) only in the limited
case of states with initial zero angular momentum
{\cite{ostrovsky_96b,ostrovsky_96c}}.

In this work we show that the accuracy of the classical results
does indeed rest on a particularly direct connection between 
classical and quantum predictions,
and demonstrate explicitly that in the perturbative limit 
the quantum expectation values of the 
angular momentum and the Runge-Lenz vector obey
{\em exactly the same equations as the time averaged classical variables}.
We also investigate the conditions on the fields
for which the perturbative approach holds, and we find that classical
mechanics seems to require more relaxed constraints on the external fields 
than quantum mechanics, as the classical condition for the conservation
of the Kepler energy -see below- is not equivalent to the quantum condition
for negligible intermanifold mixing. 
However, by studying in detail the contributions to the dynamics 
which stem from $n$-mixing, we demonstrate that the
perturbative equations of motion for the quantum expectation values
remain accurate also under the less restrictive classical conditions,
as long as the dynamics is time averaged over a Kepler period. 
In fact, the time averaged equations describe well the dynamics of 
quantum expectation values also when the Rydberg electron is
initially excited in a superposition of hydrogenic $n$-manifolds,
i.e., its initial state is not a stationary eigenstate of the unperturbed
Hamiltonian, but a time dependent wave packet.
Interestingly, time averaging is precisely the same procedure which
leads to the classical perturbative equations, and therefore our 
result is an explicit example of how the scrambling of the 
principal quantum number (quantum decoherence) 
brings about a more direct quantum-classical correspondence.
Finally, we show that for special superpositions of the eigenstates of the 
bare Hamiltonian (that is, elliptic states 
{\cite{delande_89a,delande_90a,nauenberg_89a}}) 
and in the limit of large principal quantum numbers, the quantum expectation 
values also have the appropriate quasiclassical initial conditions. 
Most importantly, elliptic states are not merely theoretical
constructs: they have been prepared in the laboratory and some of their
properties have already been studied experimentally
{\cite{ehrenreich_94a,ehrenreich_94b,ehrenreich_97a,horsdal_98a}}.

Our findings
are not merely an application of Ehrenfest's theorem {\cite{sakurai_85}}. 
Although Ehrenfest's theorem relates the time evolution 
of the quantum expectation values
to the classical equations of motion, it does not state that quantum expectation
values and classical variables obey {\em exactly} the same equations. 
Such an identity holds only for the harmonic oscillator and, albeit
trivially, also for the free particle and any linear potential. 
Because the 
harmonic oscillator is often used as a 
textbook example, it often leads to the incorrect
impression that such exact equivalence is of more general nature. 
In truth this correspondence is 
a very special property of potentials which are at most
quadratic, because in general
the expectation value of the ``force operator" 
$f( {\bf {\hat r} } ) = - \nabla V ( {\bf {\hat r} } )$,
which is a function of quantum observables,
is not equal to the same function evaluated at the expectation values
of the observables; that is (in general):
\begin{equation}
\label{q_0}
	\langle \psi | f ( {\bf {\hat r}} ) | \psi \rangle  \neq
	f ( \langle \psi | {\bf {\hat r}} | \psi \rangle ) \; .
\end{equation}
In the very special case of the harmonic oscillator, for example,
the restoring force of the spring
is simply proportional to ${\hat x}$ and the two 
sides of Eq. (\ref{q_0}) are identical,
hence the exact correspondence between quantum and classical evolution.
Also, the perturbative treatment of the Kepler problem
borrows heavily from the methods of
celestial mechanics {\cite{szebehely_67}} and studies
the dynamics of time averages of the 
classical variables. Such an approach is not the same as an
expansion of the Hamiltonian around
an equilibrium point and up to quadratic terms in the potential, 
which would make the system trivially equivalent to a harmonic oscillator.
Therefore, our work amounts to an extension of Ehrenfest theorem, in 
a much stronger form, for the important case 
of the hydrogen atom in weak, slowly varying external fields.

This paper is organized as follows: in section II we derive explicitly the 
equations of motion for the quantum expectation values over 
states confined within a hydrogenic $n$-manifold, and show that they
coincide with the classical perturbative equations to first order in the fields.
In section III we investigate the contributions to the dynamics
due to the intermanifold mixing: we show that the same 
perturbative equations of motion remain accurate even if the 
state is not initially confined within a specific $n$-manifold, 
as long as one considers the time average (over a Kepler period)
of the dynamics.
In section IV we study the initial conditions for the 
quantum expectation values over different 
quantum states and also discuss a few physical implications of our results.
Finally in section V we draw some general conclusions.

\section{Classical and Quantum Equations}

In atomic units (which we use throughout this paper) 
the Hamiltonian for a hydrogen atom in crossed electric and magnetic field
is:
\begin{equation}
\label{q_1}
	H = \frac{p^2}{2} - \frac{1}{r} - \omega_L L_{z} + Fx + \frac{ \omega_L ^2 }{2}
	( x^2 + y^2 ) \; ,
\end{equation}
where the electric field is parallel
to the $x$-axis and its 
strength is $F$; the magnetic field is antiparallel to the $z$-axis 
and $\omega_L$ is the Larmor frequency of the magnetic field, which in atomic units 
is equal to half the strength of the field. 
For weak fields the diamagnetic term, which is
proportional to the square of the field, can be neglected,
and the Hamiltonian becomes identical to the one for a hydrogen atom in 
a weak electric field rotating 
with frequency $\omega_L$, 
in the noninertial frame which rotates with the field 
{\cite{demkov_74,myself_6,myself_7,myself_8,myself_9}}.

The classical perturbative
treatment of the hydrogen atom in weak, external fields is based on 
the methods of celestial mechanics, and one is interested in the secular variation
of the elements of the Kepler ellipse followed by the classical 
electron {\cite{szebehely_67}}.
While a Kepler ellipse can be described by many equivalent sets of elements, 
the equations of motion are particularly simple if one
chooses the angular momentum and the Runge-Lenz vector.
Therefore, the dynamical variables of the classical problem are 
the time averages over a Kepler period, and along a Kepler ellipse
(which is the classical solution to zeroth order in the external fields), of
the angular momentum {\bf L} of the electron, and its 
scaled Runge-Lenz vector {\bf a}, which for bound
states is defined as {\cite{goldstein_80}}:
\begin{equation}
	\label{q_2}
	{\bf a} = \frac{1}{\sqrt{-2E}} 
        	\left\{   
                       \frac{1}{2} 
                       \left(    
                             {\bf p} \times {\bf L} - {\bf L} \times {\bf p} 
                       \right) 
                       - \frac{ {\bf r } }{r} 
                \right\} \; ,
\end{equation}
where $E = - 1 / 2 n^{2}$ is the Kepler energy of the electron.
The antisymmetrization of the cross product is not 
necessary in classical mechanics, but is
essential in quantum mechanics to obtain a hermitian operator {\cite{pauli_26a,englefield_72}}.
To first order in the external fields the classical, time averaged
angular momentum and scaled Runge-Lenz vector (which for the sake of a simpler
notation we will still indicate respectively as {\bf L} and {\bf a}) satisfy the 
following equations of motion
{\cite{born_60,hezel_92a,hezel_92b,percival_79,myself_7}}:
\begin{eqnarray}
\label{q_3}
	& & \frac{d {\bf L} }{dt} = - {\boldsymbol \omega}_{S} \times {\bf a} + 
            {\boldsymbol \omega}_{L} \times {\bf L} \nonumber \\
	& & \frac{d {\bf a} }{dt} = - {\boldsymbol \omega}_{S} \times {\bf L} +
            {\boldsymbol \omega}_{L} \times {\bf a}  \; ,
\end{eqnarray}
where ${\boldsymbol \omega}_S$ is the Stark frequency of the electric field, and is defined as:
\begin{equation}
\label{q_4}
	{\boldsymbol \omega}_S = \frac{3}{2} n {\bf F} \; ,
\end{equation}
and ${\boldsymbol \omega}_L$ is the Larmor frequency vector: 
it is directed along the external magnetic field and
its magnitude is equal to the Larmor frequency of the field itself.

Eqs. (\ref{q_3}) were derived originally 
by Born {\cite{born_60}} and in classical mechanics
they remain accurate as long as the 
two frequencies (Stark and Larmor) are much smaller 
than the Kepler frequency $\omega_K$ of the electron
{\cite{born_60,hezel_92a,hezel_92b,percival_79,myself_7}}:
\begin{equation}
\label{q_5}
	\omega_S \, ,  \, \omega_L \ll \omega_K = \frac{1}{n^3} \; .
\end{equation}
In classical mechanics Eq. (\ref{q_5}) means that the elements of the Kepler orbit
do not vary significantly over a Kepler period, so that 
the classical electron still moves, to a good approximation, along a Kepler ellipse,
and the Kepler energy of the classical electron is conserved. In fact, 
the classical angular momentum and the Runge-Lenz vector obey two constraint 
equations {\cite{goldstein_80}}:
\begin{eqnarray}
\label{q_5a}
	& & {\bf L} \cdot {\bf a} = 0 \nonumber \\
	& & L^{2} + a^{2} = n^{2} \; .
\end{eqnarray}
It is easy to see that both constraints are invariant under the time 
evolution dictated by Eqs. (\ref{q_3}), and also that the second of Eqs. (\ref{q_5a})
implies the conservation of the Kepler energy of the electron.

Instead, the quantum mechanical interpretation is that the 
external fields remove the degeneracy of the 
$n^{2}$ unperturbed states of the hydrogenic $n$-manifold,
and the conditions of Eq. (\ref{q_5}) mean that the 
energy difference between two adjacent perturbed states 
is much smaller than the separation between 
adjacent, unperturbed Rydberg energy levels.
However, this is not the usual condition under which 
in quantum mechanics
inter-{\em n} mixing is negligible. 
For example, in the case of just an external d.c. field (the extension to 
include also a magnetic field is straightforward {\cite{turgay_96b}})
the energy separation between the lowest
and the highest Stark states for a fixed principal quantum number {\em n}
is (to first order in the field) {\cite{bethe_77}}:
\begin{equation}
\label{q_5b}
	\Delta E = 3 n ( n - 1 ) F \; .
\end{equation}
Therefore the approximate condition for level crossing of the highest
Stark state from a given {\em n}-manifold with the lowest Stark 
level from the next {\em n}-manifold is given by the Inglis-Teller limit
{\cite{gallagher_94}}:
\begin{equation}
\label{q_5c}
	3 n^{2} F \approx \frac{1}{n^{3}} \; .
\end{equation}
Clearly, in the semiclassical limit
the quantum condition on the external fields
for negligible $n$-mixing of Eq. (\ref{q_5c}) is
much stronger than the classical condition of Eq. (\ref{q_5}).
We show below, however, that 
the perturbative treatment of the dynamics
of the quantum expectation values
remains accurate also in presence of some
degree of $n$-mixing induced by the external fields, as long as the 
dynamics is time averaged over a Kepler period.

In this section  we confine our
study to the dynamics of the quantum expectation values 
of the angular momentum and the scaled 
Runge-Lenz vector operators 
(that is, $\langle \psi | {\bf {\hat L}} | \psi \rangle$ and 
$\langle \psi | {\bf {\hat a}} | \psi \rangle$; throughout
this paper we use boldface letters for vectors, and
a caret indicates a quantum operator, not a unit vector 
which we denote instead as {\bf e}$_{\imath}$) 
over superpositions of the hydrogen atom eigenfunctions with a well defined
principal quantum number; i.e., 
over states $| \psi_{n} \rangle$ 
which are localized within a hydrogenic $n$-manifold.
More precisely, we show here that to first order in the external fields
the quantum expectation values obey exactly the same 
equations of motion as Eqs. (\ref{q_3}). 

The cornerstone of the study of hydrogenic systems in weak, external fields
is the so-called Pauli's repacement
{\cite{pauli_26a,englefield_72}},
according to which the matrix elements 
{\em between states within the same n-manifold} 
of the position operator are directly proportional to 
the corresponding matrix elements of the Runge-Lenz vector operator:
\begin{equation}
\label{lallaballa}
	\langle n  \ell^{\prime} m^{\prime} | {\hat r}_{\imath} | n \ell m \rangle =
        - \frac{3}{2} n
        \langle n  \ell^{\prime} m^{\prime} | {\hat a}_{\imath} | n \ell m \rangle  \; .
\end{equation}
By replacing the position operator, which appears in the perturbation Hamiltonian
for an external field, with $-3 n {\hat {\bf a }} / 2$
the demonstration of quantum-classical equivalence
{\em within an n-manifold} is straightforward 
{\cite{nauenberg_94a,carlos_97c,carlos_98a}},
but nothing can be said about the intermanifold dynamics.
Instead, in our argument we do not apply Pauli's replacement
directly in the Hamiltonian; our approach is more
complicated, but it makes possible the extension 
of our analysis (in the next section)
also to the dynamics of Rydberg wave packets.
Moreover, we will be able to show that the quantum analogs
of Eqs. (\ref{q_3}) remain accurate under the more relaxed, 
classical conditions 
on the external fields given in Eq. (\ref{q_5}).

To prove the special quantum-classical equivalence, 
we will make use of the following identity, which holds within 
a hydrogenic manifold and which we derive explicitly in 
Appendix A:
\begin{equation}
\label{q_6}
	\langle \psi_{n} | {\hat r}_{\imath} {\hat p}_{\jmath} | \psi_{n} \rangle = 
	- \langle \psi_{n} | {\hat p}_{\imath} {\hat r}_{\jmath} | \psi_{n} \rangle \; ,
\end{equation}
where ${\hat r}_{\imath}$ and ${\hat p}_{\jmath}$ are 
components of the position and momentum operator respectively. 

Indeed, armed with the result of Eq. {\ref{q_6} it is easy to show that to first order in the 
external fields the expectation values of the quantum observables satisfy the same
equations as the time averages (over a Kepler period) of the classical variables.

The equations of motion of the quantum expectation values are straightforward in the 
Heisenberg picture {\cite{sakurai_85}}:
\begin{eqnarray}
\label{q_18}
	& & \langle \psi_{n} | \frac{ d {\hat L}_{\imath} }{ dt } | \psi_{n} \rangle =
           - i \langle \psi_{n} | \left[ {\hat L}_{\imath} , 
	   {\hat H} \right] | \psi_{n} \rangle 
           \nonumber \\
	& & \langle \psi_{n} | \frac{ d {\hat a}_{\imath} }{ dt } | \psi_{n} \rangle =
           - i \langle \psi_{n} | \left[ {\hat a}_{\imath} , 
	   {\hat H} \right] | \psi_{n} \rangle \; .
\end{eqnarray}
and we now show that they are identical to Eqs. (\ref{q_3}). 

The classical equations (\ref{q_3}) contain two terms, an electric term which is
proportional to the Stark frequency $\omega_S$, and which couples the angular momentum 
to the scaled Runge-Lenz vector; and a magnetic term which is proportional to 
the Larmor frequency $\omega_L$ of the field
(or to the rotation frequency of a slowly rotating electric field, in a noninertial frame
rotating with the field itself
{\cite{demkov_74,myself_6,myself_7,myself_8,myself_9}}).
Both the scaled Runge-Lenz vector and the angular momentum commute with the 
hydrogenic Hamiltonian (they are invariants of the pure Kepler problem) {\cite{goldstein_80}}.
Moreover, it is easy to see that the magnetic term of the classical equations
can be recovered by invoking the vectorial properties of
${\bf {\hat a}}$ and ${\bf {\hat L}}$, because of which
their commutators with the magnetic term in of the Hamiltonian
(i.e., $- \omega_L {\hat L}_z$)
obey the well known rule {\cite{sakurai_85}}: 
\begin{equation}
\label{eq_5_18a}
	\left[ {\hat V}_{\imath} , {\hat L}_{\jmath} \right] = 
	i \epsilon _{\imath, \jmath, k} {\hat V}_{k} \; ,
\end{equation}
where ${\hat V}_{\imath}$ stands for the $\imath^{th}$ component of any vector operator.

Therefore, we only need to investigate 
the commutators of ${\bf {\hat a}}$
and ${\bf {\hat L}}$ with the electric perturbation $F {\hat x}$.

We begin with ${\hat a}_{y}$: 
\begin{equation}
\label{q_19}
	\begin{split}
	-i F 
	& 
	\left[ {\hat a}_{y} , {\hat x} \right]
	= -i \frac{n}{2} F 
	\left\{
	\left[ 
	\left( 
	{\hat p}_z {\hat L}_x - 
        {\hat p}_x {\hat L}_z 
	\right) ,
	{\hat x} 
	\right] \right. \\
	&
	\left.
	- \left[ 
	\left( 
	{\hat L}_z {\hat p}_x - 
	{\hat L}_x {\hat p}_z 
	\right) ,
        {\hat x} 
        \right]     
	\right\} \! = 
	- i \frac{n}{2} F \!
        \left\{
       	-{\hat p}_x \left[ {\hat L}_z , {\hat x} \right] 
	\right. \\
	&
	\left.
        - \left[ {\hat p}_x , {\hat x} \right] {\hat L}_z -
        {\hat L}_z \left[ {\hat p}_x , {\hat x} \right] - 
        \left[ {\hat L}_z , {\hat x} \right] {\hat p}_x 
        \right\} \\
	&
	= n F \left\{ {\hat L}_z - {\hat y} {\hat p}_x  \right\} \; .
       	\end{split}
\end{equation}
However, using the identity of Eq. (\ref{q_6}) one has:
\begin{equation}
\label{q_20}
	- \langle \psi_{n} |
        {\hat y} {\hat p}_x 
        | \psi_{n} \rangle =
        \frac{1}{2} \langle \psi_{n} |
        {\hat x} {\hat p}_y - {\hat y} {\hat p}_x 
        | \psi_{n} \rangle \; ,
\end{equation}
from which it follows:
\begin{equation}
\label{q_21}
	- i F \langle \psi_{n} | \left[ {\hat a}_{y} , 
	{\hat x} \right] | \psi_{n} \rangle
        = \frac{3}{2} n F \langle \psi_{n} | {\hat L}_{z} 
	| \psi_{n} \rangle \; .\
\end{equation}
This is the same as the electric term in the equation of motion 
for the classical time averaged $a_y$. 

The derivation of the electric term for the equation of
of $\langle \psi_{n} | {\hat a}_z | \psi_{n} \rangle$
follows along the same lines and it is easy to see that it
yields the desired result.

We then consider ${\hat a}_x$:
\begin{equation}
\label{q_22}
	\begin{split}
	- i F 
        & 
	\left[ {\hat a}_{x} , {\hat x} \right] = 
	\frac{n}{2} F 
        \left\{
        \left[ 
        \left( 
        {\hat p}_y {\hat L}_z - 
        {\hat p}_z {\hat L}_y 
        \right) ,
        {\hat x} 
        \right] \right. \\
	&
	- \left.
        \left[ 
        \left( 
        {\hat L}_y {\hat p}_z - 
        {\hat L}_z {\hat p}_y 
        \right) ,
        {\hat x} 
        \right]     
        \right\} = 
	- i \frac{n}{2} F 
        \left\{
        {\hat p}_y \left[ {\hat L}_z , {\hat x} \right] \right. \\
	&
	-
	\left.
        {\hat p}_z  \left[ {\hat L}_y , {\hat x} \right] -
        \left[ {\hat L}_y , {\hat x} \right] {\hat p}_z  + 
        \left[ {\hat L}_z , {\hat x} \right] {\hat p}_y 
        \right\} \\
        & 
	= \frac{n}{2} F 
        \left\{
        {\hat p}_y {\hat y} + {\hat y} {\hat p}_y +
        {\hat p}_z {\hat z} + {\hat z} {\hat p}_z 
        \right\} 	\; .
        \end{split}
\end{equation}
Invoking once again Eq. (\ref{q_6}) one obtains immediately:
\begin{equation}
       \label{q_23}
       F \langle \psi_{n} | \left[ {\hat a}_{x} , {\hat x} \right] | \psi_{n} \rangle
       = 0	\; ,
\end{equation}
which is the same as the right hand side of the corresponding classical equation
of motion.

Finally, we turn to the equations for the angular momentum.
The classical equations can be written as:
\begin{equation}
\label{q_23a}
	\frac{ d L_{\imath} }{ dt } =
	- \epsilon_{\imath ,\jmath ,k} {\omega_{S}}_{\jmath} a_{k} = 
	- \frac{3}{2} n F \epsilon_{\imath ,1,k} a_{k} \; ,
\end{equation}
where we have specialized the right hand side to the case of an 
external field along the $x$-axis.
Using once again the 
vector properties of ${\bf {\hat r}}$, the contribution of the 
electric term to the quantum equations is:
\begin{equation}
\label{q_24}
	- i F \left[ {\hat L}_{\imath} , {\hat x} \right] = 
        F \epsilon_{\imath ,1,k} {\hat r}_k	\; .
\end{equation}
This is not yet in the desired form. However, within a given
$n$-manifold one can apply 
Pauli's replacement {\cite{pauli_26a,englefield_72}}
according to which:
\begin{equation}
\label{q_25}
	\langle n  \ell^{\prime} m^{\prime} | {\hat r}_{\imath} | n \ell m \rangle =
        - \frac{3}{2} n
        \langle n  \ell^{\prime} m^{\prime} | {\hat a}_{\imath} | n \ell m \rangle  \; .
\end{equation}
Pauli's replacement is mathematically exact, and yet physically it is just 
an approxximation, because the dynamics of the electron is 
only approximately confined within a given n-manifold.
Clearly, the accuracy of the approximation rests
on certain conditions on the external fields, depending on which the dynamics
may -or may not- be very well localized within a hydrogenic manifold, and we discuss such 
conditions in detail below.
However, in the present section we are interested only the intramanifold
dynamics, and therefore one has:
\begin{equation}
       \label{q_26}
       F \epsilon_{\imath ,1,k} 
       \langle \psi_{n} | {\hat r}_k | \psi_{n} \rangle  = 
       - \frac{3}{2} n F \epsilon_{\imath ,1,k} 
       \langle \psi_{n} | {\hat a}_k | \psi_{n} \rangle  \; ,
\end{equation}
which is the same electric term as in the right hand side of the corresponding
classical equations of motion,
and our proof is complete.

Our derivation of the equations of motion of the quantum expectation values
is accurate only to first order in the fields because it
relies heavily on the identity of Eq. (\ref{q_6}), which holds for the 
unperturbed $| \psi_{n} \rangle$ states; these are eigenstates of the 
hydrogen atom Hamiltonian and therefore
are {\em quantum solutions to zeroth order in the fields}.
In fact, this is exactly in the same spirit as the classical
approach, where the right hand sides of Hamilton equations are
time averaged over a Kepler period and, most importantly, 
along Kepler ellipses, which are the
{\em classical solutions to zeroth order in the fields}.

The same consideration can also be cast in the language of operators, by
observing that we have proven Eq. (\ref{q_6}) only for the
time independent operators of the Schr{\"o}dinger picture, whereas
in the right hand side of the equations of motion 
one must more correctly use the time dependent operators of the Heisenberg picture.
However, in a first order approximation one may assume that
the time evolution of the 
operators in the Heisenberg picture is dictated solely by the
hydrogenic propagator, which commutes with both ${\bf {\hat L}}$ and ${\bf {\hat a}}$,
so that one can legitimately use the properties of 
those operators in the Schr{\"o}dinger picture.

Since Eq. (\ref{q_6}) holds for all possible pairs of indexes $\{ \jmath,k \}$,
the same derivation can be easily extended to the case of 
slowly varying (both in magnitude and direction) electric and magnetic fields,
in which case the perturbing Hamiltonian ${\hat H}_{1}$ becomes:
\begin{equation}
	\label{q_27}
	{\hat H}_{1} = 
        \sum_{\imath} \left\{ F_{\imath} (t) {\hat r_{\imath} } - 
	{\omega_{L}}_{\imath} (t) {\hat L_{\imath} } \right\} \; .
\end{equation}

However, the most important feature of 
our proof is that we have not applied Pauli's replacement
directly in the Hamiltonian
{\cite{nauenberg_94a,carlos_97c,carlos_98a}}.
An early application of Pauli's replacement 
yields a straightforward proof of quantum-classical equivalence 
{\em within an n-manifold}, but it
erases all information about
the precise conditions on the fields under which the perturbative classical
equations of motion constitute an accurate description of the 
dynamics of the quantum expectation values. Moreover, it
also makes impossible to study the corrections to the dynamics
due to intermanifold mixing and therefore one could not extend
Eqs. (\ref{q_3}) to the case of Rydberg wave packets.
Instead, in the next section 
we address in detail precisely these important issues.

\section{intermanifold dynamics and quantum-classical correspondence in Rydberg wave packets}

In this section we study intermanifold mixing and 
the conditions on the external fields under which the classical perturbative
equations of motion offer an accurate treatment of the dynamics of the 
quantum expectation values. We will show that the same conditions as in 
classical mechanics hold in quantum mechanics too, provided that the dynamics is
time averaged over a Kepler period. Most importantly, we will demonstrate that upon
time averaging Eqs. (\ref{q_3}) apply also
to the case of Rydberg wave packets.

For the sake of simplicity we restrict our
analysis to the pure Stark case, that is when there is no external magnetic
field; the extension to the more general case including a weak 
magnetic field is straightforward. 
Therefore, in this section we assume a simplified Hamiltonian:
\begin{equation}
\label{q_28}
	{\hat H} = {\hat H}_{0} + F {\hat x} \; ,
\end{equation}
where ${\hat H}_{0}$ is the hydrogen atom Hamiltonian and $F$ is again the 
strength of the external electric field.

Inter-manifold mixing is due to two main causes, depending on how the 
Rydberg state is prepared. 
If the Rydberg electron is initially 
confined within a hydrogenic $n$-manifold, then $n$-mixing is induced 
by the applied external field, and in that case the intermanifold
contributions to the equations of motion 
are of second order in the applied field. 
Note that this is a very realistic picture for slow ion-Rydberg collisions
{\cite{macadam_81,macadam_85,macadam_88a,macadam_93,macadam_95,myself_7,myself_9}}.
In fact, ion-Rydberg collisions 
are actually gentle encounters at very large ion-Rydberg separation, which are 
effective because of the long-range nature of the coulomb interaction, and are 
very accurately modeled by a time dependent, weak external field acting on the 
Rydberg electron {\cite{myself_7,myself_9}}.
Typically, the Rydberg state is prepared in absence of
external dc fields, and the weak field of the colliding ion is turned on
adiabatically as the ion slowly approaches the Rydberg atom.
In such situation, the analysis of intermanifold mixing 
is equivalent to the study of the second order corrections to the equations of motion,
and it allows one to determine for what precise conditions on the 
external electric field such corrections are negligible. 

However, $n$-mixing can also be present at the outset, either if the 
Rydberg state is prepared in presence of the applied dc field and the field
is strong enough to mix adjacent Rydberg levels,
or alternatively if a short, large
bandwidth laser pulse is employed in the preparation of the 
Rydberg state. In both cases
the ground state is coupled to a {\em distribution} of hydrogenic
manifolds, and the Rydberg electron is not excited to a high energy, 
stationary eigenstate of the hydrogen atom, but rather to some 
time dependent wave packet. 
To a first approximation the Rydberg wave packet
oscillates with the Kepler frequency of the eigenstate around which
the distribution of principal quantum numbers
is centered; therefore a Rydberg wave packet contributes 
rapidly oscillating,
intermanifold terms to the equations of motion 
for the quantum expectation values. Such intermanifold
contributions may be of first order in the external field. 
In this section we extend the validity of the classical,
perturbative Eqs. (\ref{q_3}) precisely to the case of
Rydberg wave packets, by time averaging the equations of motion over 
a Kepler period and by showing that the {\em secular}, 
intermanifold contributions to the 
dynamics remain negligible under the classical conditions 
of Eq. (\ref{q_5}) for the external field.

More precisely, the wave function of a Rydberg wave packet is:
\begin{equation}
\label{q_29}
	| \psi (t) \rangle = \sum_{n} C_{n} | \psi_{n} (t) \rangle \; ,
\end{equation}
and the Heisenberg equations of motion for the expectation values of 
either ${\bf {\hat L}}$ or ${\bf {\hat a}}$ over the state of Eq. (\ref{q_29})
include ``off-diagonal" matrix elements
(for the sake of brevity we
call ``off-diagonal" the matrix elements of any operator
between states from two different hydrogenic manifolds; 
whereas we will call ``diagonal" the matrix elements 
between two states within the same $n$-manifold, regardless of 
their angular momentum quantum numbers).

For example, if we indicate generically by 
${\hat O}_{cl} + {\hat O}_{q}$ 
the combinations of operators in the right hand side of 
the Heisenberg equations of motion,
in the case of ${\hat L}_{\imath}$ one has:
\begin{equation}
\label{q_30}
	\begin{split}
	\frac{ d }{ dt }
	\langle \psi | 
	&
	{\hat L}_{\imath} (t)
	| \psi \rangle
	=
	F \Big{\{} 
	\sum_{n} 
	\left| C_{n} \right| ^{2} 
	\langle \psi_{n} | 
	{\hat O}_{cl} (t) + {\hat O}_{q} (t)
	| \psi_{n} \rangle 
%
%%%%%%%%%%%%%%%%%%%%%%%%%%%%%%%%%%%%%%%%%%%%%%%%%%%%%%%%%%%%%%%%%%%%%%%%%%%%%%%%%%%%%%%%%%%%%%
%
% This is done to split the goddam equation on two pages
% Comment out when preparing the preprint version
%
\end{split}
\end{equation}
%%\pagebreak
{\vspace*{-1.1cm}}
\begin{equation}
\begin{split}
\nonumber
%
%
%%%%%%%%%%%%%%%%%%%%%%%%%%%%%%%%%%%%%%%%%%%%%%%%%%%%%%%%%%%%%%%%%%%%%%%%%%%%%%%%%%%%%%%%%%%%%
%
\\
	& +
	\sum 
	_{ \scriptstyle{ n^{\prime}, n } \atop \scriptstyle{  \ n^{\prime} \neq n } }
	{\bar C}_{n ^{\prime} } 
	C_{n}
	\langle \psi_{n ^{\prime} } | 
	{\hat O}_{cl} (t) + {\hat O}_{q} (t)
	| \psi_{n} \rangle \Big{\}} \; .
	\end{split}
\end{equation}
where ${\hat O}_{cl}$ indicates a combination of operators which 
correspond to the classical variables
in the right hand side of the classical equations of motion (i.e., 
a combination of angular momentum and Runge Lenz vector). 
Instead, ${\hat O}_{q}$ indicates the purely quantum corrections, 
which vanish when the motion is exactly confined within a
hydrogenic manifold.
Obviously, an identical expression holds also for the equation of motion of the 
expectation value of ${\hat a}_{\imath}$

The expectation values of the operators of
${\hat O}_{cl}$ appear in the equations of motion in the same 
way as the classical time averaged variables,
and therefore they evolve in time exactly like their classical 
counterparts. This is true even if the state is not confined 
within a hydrogenic manifold. 
Yet, for states which are spread
over more than one $n$-manifold the quantum contributions
from the expectation value of ${\hat O}_{q}$ do not vanish exactly.
In fact, we proved in the previous section and in Appendix A
that only the diagonal matrix elements of ${\hat O}_{q}$ vanish,
but the same does not hold for the off-diagonal matrix elements.

However, we show below that under the classical conditions for
the external field, and upon time averaging, 
all the off-diagonal matrix elements of the double
sum of Eq. (\ref{q_30}) offer a negligible contribution to the dynamics.
In our demonstration we do not distinguish between the two operators, and treat
${\hat O}_{cl} + {\hat O}_{q}$ as a single term which,
for the sake of brevity, we simply denote as ${\hat O}$.
That is, our argument demonstrates that also the off-diagonal
matrix elements of ${\hat O}_{cl}$ yield only
negligible contributions to the equations 
of motion. Therefore, the quantum dynamics of the
expectation values of the operators of ${\hat O}_{cl}$ is determined only by 
the intramanifold terms. Note that this last observation is not essential to 
the issue of quantum-classical correspondence.

Before our demonstration, however, we must discuss briefly the operators of 
Eq. (\ref{q_30}) and the magnitude of their
matrix elements between two eigenstates of the hydrogen atom.
From the previous section, and also from Appendix A,
it is easy to see that the sum of the two combination of operators 
${\hat O}_{cl} + {\hat O}_{q} = {\hat O}$
may be equal to one or a combination of the following operators:
\begin{equation}
\label{q_31}
	{\hat O} \sim
	\begin{cases}
		{\hat r}_{\imath} &	\\
		n  {\hat p}_{\imath} {\hat r}_{\jmath} & \imath,\jmath = 1,2,3 \\
		n {\hat L}_{\imath} & 
	\end{cases} \; ,
\end{equation}
First, the $n {\hat L}_{\imath}$ operator yields matrix elements
the magnitudes of which are at most $ \lesssim n^{2} $.
Moreover, ${\hat L}_{\imath}$ 
commutes with the hydrogen atom Hamiltonian
and therefore all its off-diagonal matrix elements vanish, and 
it is very easy to prove that its intermanifold contributions 
to the dynamics are negligible (see below).

Next, in the case of ${\hat r}_{\imath}$ 
the magnitude of the matrix elements and, most importantly, their
{\em scaling} with $n$ are determined solely by the {\em radial} matrix elements:
\begin{equation}
\label{q_31a}
	R_{ n , \ell }^{ n^{\prime} , \ell^{\prime} } 
	= \langle n \ell m | {\hat r} | n^{\prime} \ell^{\prime} m^{\prime} \rangle \; .
\end{equation}
If $n^{\prime} = n$ one has {\cite{bethe_77}}:
\begin{equation}
\label{q_31b}
	R_{ n , \ell-1 }^{ n , \ell } =
	R_{ n , \ell }^{ n , \ell-1 } = -
	\frac{3}{2} n^{2} \sqrt{ 1 - \frac{ \ell^{2} }{ n^{2} } } \; ,
\end{equation}
whereas if $n^{\prime} \neq n$ the radial matrix elements of the position operator
are given by a complicated formula which 
involves hypergeometric functions {\cite{bethe_77}}.
However for $n^{\prime} , n \gg 1 $,
$\left| R_{ n , \ell }^{ n^{\prime} , \ell^{\prime} } \right|$
is accurately approximated
by a well known semiclassical result {\cite{percival_75a}}:
\begin{equation}
\label{q_32}
	\begin{split}
	&
	\left|  
	R_{ n , \ell }^{ n^{\prime} , \ell^{\prime} } 
	\right| \\
	& \approx
	\left| \frac{ n_{c}^{2} }{ 2 \Delta_{ n , n^{\prime} } } 
	\left\{
	\left(
	1 - \Delta_{ \ell , \ell^{\prime} } \frac{ \ell_{c} }{ n_{c} }
	\right)
	J_{ ( \Delta_{ n , n^{\prime} } + 1 )}
	\left( \Delta_{ n , n^{\prime} } \epsilon \right ) 
	\right. \right. \\
	&
	\left. \left.
	-
	\left(
	1 + \Delta_{ \ell , \ell^{\prime} } \frac{ \ell_{c} }{ n_{c} }
	\right)
	J_{ ( \Delta_{ n , n^{\prime} } -1 ) }
	\left( \Delta_{ n , n^{\prime} } \epsilon \right )
	\right\} \right|
	\lesssim 
	\frac{ n^{2} }{ 2 \left| \Delta_{ n , n^{\prime} } \right| }
	\; ,
	\end{split}
\end{equation}
where $\Delta_{ n , n^{\prime} } = n - n^{\prime}$ and 
$\Delta_{ \ell , \ell^{\prime} }  = \ell - \ell^{\prime}$. 
In Eq. (\ref{q_32})
we also used:
$n_{c} = 2 n n^{\prime} / ( n + n^{\prime} )$,
$\ell_{c} = {\rm max}(\ell, \ell^{\prime})$ and finally 
$\epsilon ^{2} = 1 - \ell_{c}^{2} / n^{2}$.
Clearly the last inequality of Eq. (\ref{q_32}) is accurate only to the leading 
order in $n$, and to the same order
it is equally correct if one uses either $n$ or $n^{\prime}$ or $n_{c}$. 

Finally, for $n {\hat p}_{\imath} {\hat r}_{\jmath}$ by using 
${\hat p}_{\imath} = -i \left[ {\hat r}_{\imath} , {\hat H}_{0} \right]$
and inserting a resolution of unity between the two operators one has:
\begin{equation}
\label{q_33}
	\begin{split}
	\left|
	n
	\right.
	&
	\left.
	\langle n^{\prime} \ell^{\prime} m^{\prime} | 
	{\hat p}_{\imath} {\hat r}_{\jmath} 
	| n^{\prime \prime} \ell^{\prime \prime} m^{\prime \prime} \rangle 
	\right| 
	\\
	&
	\approx 
	\left|
	i n
	\sum_{ \mu }
	\frac{ \Delta_{ n^{\prime} , \mu } }{ \mu {n^{\prime}}^{2} }
	\langle n^{\prime} \ell^{\prime} m^{\prime} | 
	{\hat r}_{\imath} 
	| \mu \rangle \langle \mu |
	{\hat r}_{\jmath} 
	| n^{\prime \prime} \ell^{\prime \prime} m^{\prime \prime} \rangle \right| < \\
	&
	<  
	\left|
	\langle n^{\prime} \ell^{\prime} m^{\prime} | 
	{\hat r}_{\jmath}
	| n^{\prime \prime} \ell^{\prime \prime} m^{\prime \prime} \rangle \right| \; ,
	\end{split}
\end{equation}
where $\Delta_{ n^{\prime} , \mu } = n^{\prime} - n_{\mu}$, and where
we used the final result of Eq. (\ref{q_32}).
In Eq. (\ref{q_33})
for the sake of a simpler notation we adopted the following convention:
$ | \mu \rangle  = | n_{\mu} \ell_{\mu} m_{\mu} \rangle $ (which we will often
use in this section).

The result of Eq. (\ref{q_33}) rests on the observation that in the 
semiclassical limit
the radial matrix elements of ${\hat r}_{\imath}$ 
become rapidly very small for large $\Delta$'s, 
as one can easily see from Eq. (\ref{q_32});
and one may safely assume that for 
nonnegligible matrix elements the difference between the two
principal quantum numbers is always much smaller than any
of the principal quantum numbers themselves 
(e.g. $\Delta_{n^{\prime}, \mu} \ll n^{\prime} , n_{\mu} $). 
Therefore one may legitimately neglect higher order corrections
in $\Delta_{ n^{\prime}, {\mu} } / n^{\prime}$.

In fact, we assume precisely this important condition throughout our argument: 
that is, we assume that the variance 
of the distribution of the Rydberg wave packets over the 
hydrogenic principal quantum numbers 
is always much smaller than the principal quantum number at the center of 
the distribution, i.e., the approximate 
average principal quantum number of the 
Rydberg wave packet. 
This is a very realistic 
approximation for most laser pulses employed in the excitation of Rydberg electrons,
and it breaks down only for ultrashort, ultralarge bandwidth pulses; or
when the Rydberg state is excited in presence of ultrastrong external fields.

In what follows we conduct our analysis in the most general form. 
Although 
at some point we 
specialize our argument to the case
${\hat O} = {\hat r}_{\imath}$ which yields the largest off-diagonal
contribution to the equations of motion, it will be easy to see that
the treatment of the case
${\hat O} = n {\hat p}_{\imath} {\hat r}_{\jmath}$ 
is completely analogous.

Any off-diagonal matrix element of Eq. (\ref{q_30}) can be written as:
\begin{equation}
\label{q_34}
	\begin{split}
	\langle \psi_{ n^{\prime} } |
	{e} ^{ i {\hat H} t } {\hat O} {e} ^{ - i {\hat H} t }
	&
	| \psi_{n} \rangle = 
	\sum_{ \ell^{\prime} , m^{\prime} } 
	\sum_{ \ell , m } 
	{\bar C}_{ n^{\prime} } ( \ell^{\prime} , m^{\prime} )
	C_{n}( \ell , m) \\
	& \times
	\langle n^{\prime} \ell^{\prime} m^{\prime} | 
	{e} ^{ i {\hat H} t } {\hat O} {e} ^{ - i {\hat H} t }
	| n \ell m \rangle \; ,
	\end{split} 
\end{equation}
where we have expanded the states
$| \psi_{n^{\prime}} \rangle$ and $| \psi_{n} \rangle$, which are initially 
confined within the $n^{\prime}$- and $n$-manifold respectively, as follows:
\begin{equation}
\label{q_1200}
	| \psi_{n} \rangle =
	\sum_{ \ell , m } C_{n} ( \ell , m ) | n \ell m \rangle  \; ,
\end{equation}
where the $C_{n}( \ell , m)$'s are some general coefficients, possibly
complex. 

As we mentioned before, these off-diagonal matrix elements are present in the 
equations of motion either when the high-energy electron is prepared in
a wave packet, or when the
electron is initially confined within a single $n$-manifold, in which case
they represent the second order (in the external field) corrections to the dynamics.

The dynamics of the matrix elements of
Eq. (\ref{q_34}) is best
studied in the interaction picture {\cite{sakurai_85}}, and therefore 
it is convenient to set:
\begin{equation}
\label{q_35}
	\begin{split}
	&
	{e} ^{ - i {\hat H} t } | n \ell m \rangle =
	\sum_{\lambda} \alpha_{\lambda} (t) 
	{e}^{ - i E_{\lambda} t } | \lambda \rangle \\
	&
	\langle n^{\prime} \ell^{\prime} m^{\prime} | {e} ^{ + i {\hat H} t } =
	\sum_{\mu} \beta_{\mu} (t) 
	{e}^{ + i E_{\mu} t } \langle \mu | \; ,
	\end{split}
\end{equation}
where we used once again the convention 
$| \lambda \rangle = | n_{\lambda} \ell_{\lambda} m_{\lambda} \rangle$ and 
$| \mu \rangle = | n_{\mu} \ell_{\mu} m_{\mu} \rangle$, as 
the most important features
of our argument depend on the spectrum of the hydrogen atom, and 
are determined solely by the principal quantum number of the state.

The equations of motion for the $\alpha$'s and the $\beta$'s can be derived 
directly from the Schr{\"o}dinger equation:
\begin{equation}
\label{q_35a}
	\begin{split}
	&
	i {\dot{ \alpha }_{\lambda} } =
	F \sum_{\lambda_{1}} 
	\langle \lambda | {\hat x} | \lambda_{1} \rangle
	\alpha_{\lambda_{1}} {e} ^{ - i ( E_{\lambda_{1}} - E_{\lambda} ) t } \\
	&
	i {\dot{ \beta }_{\mu} } =
	- F \sum_{\mu_{1}} 
	\langle \mu_{1} | {\hat x} | \mu \rangle
	\beta_{\mu_{1}} {e} ^{ - i ( E_{\mu} - E_{\mu_{1}} ) t }  \; ,
	\end{split}
\end{equation}
and the solution of Eqs. (\ref{q_35a}) to 
zeroth order in the field is:
\begin{equation}
\label{q_36}
	\begin{split}
	&
	\alpha_{n_{\lambda} \ell_{\lambda} m_{\lambda} }^{ ( F^{0} ) } = 
	\delta_{n_{\lambda} , n} \delta_{\ell_{\lambda} , \ell} \delta_{m_{\lambda} , m} 
	\\
	&
	\beta_{n_{\mu} \ell_{\mu} m_{\mu} }^{ ( F^{0} ) } = 
	\delta_{n_{\mu} , n^{\prime}} \delta_{\ell_{\mu} , l^{\prime}} 
	\delta_{m_{\mu} , m^{\prime}} \; .
	\end{split}
\end{equation}
In a first order approximation Eqs. (\ref{q_35a}) become:
\begin{equation}
\label{q_37}
	\begin{split}
	&
	i {\dot{ \alpha }_{\lambda}^{(F)} } =
	F 
	\langle \lambda | {\hat x} | n \ell m \rangle
	{e} ^{ - i ( E_{ n } - E_{\lambda} ) t } \\
	&
	i {\dot{ \beta }_{\mu}^{(F)} } =
	- F 
	\langle n^{\prime} \ell^{\prime} m^{\prime} | {\hat x} | \mu \rangle
	{e} ^{ + i ( E_{n^{\prime}} - E_{\mu} ) t }  \; ,
	\end{split}
\end{equation}
and their solution is straightforward:
\begin{equation}
\label{q_38}
	{\hspace{-0.3cm}}
	\begin{cases}
	\alpha_{\lambda}^{(F)} =
	F 
	\langle \lambda | {\hat x} | n \ell m \rangle
	\frac{ {e}^{ - i ( E_{n} - E_{\lambda} ) t } - 1 }
	{ E_{n} - E_{\lambda} }
	&
	n_{\lambda} \neq n \\
	\alpha_{ \lambda }^{(F)} = - i F \langle \lambda | {\hat x} | n \ell m \rangle t 
	&
	{\hspace{-1.2cm}}
	n_{\lambda} = n, \ \{ \ell_{\lambda}, m_{\lambda} \} \neq \{ \ell , m \} \\
	\alpha_{ \lambda }^{(F)} = 1
	&
	{\hspace{-1.2cm}}
	\{ n_{\lambda}, \ell_{\lambda}, m_{\lambda} \} = \{ n , \ell , m \}
	\ ,
	\end{cases} 
\end{equation}
and also:
\begin{equation}
\label{q_38a}
	{\hspace{-.2cm}}
	\begin{cases}
	\beta_{\mu}^{(F)} =
	F 
	\langle n^{\prime} \ell^{\prime} m^{\prime} | {\hat x} | \mu \rangle
	\frac{ {e}^{ + i ( E_{n^{\prime}} - E_{\mu} ) t } - 1 }
	{ E_{n^{\prime}} - E_{\mu} }
	&
	n_{\mu} \neq n^{\prime} \\
	\beta_{ \mu }^{(F)} =  
	i F \langle n^{\prime} \ell^{\prime} m^{\prime} | {\hat x} | \mu \rangle t 
	&
	{\hspace{-1.8cm}}
	n_{\mu} = n^{\prime}, \ 
	\{ \ell_{\mu}, m_{\mu} \} \neq \{ \ell^{\prime} , m^{\prime} \} \\ 
	\beta{ \mu }^{(F)} = 1
	&
	{\hspace{-1.8cm}}
	\{ n_{\mu}, \ell_{\mu}, m_{\mu} \} = \{ n^{\prime} , \ell^{\prime} , m^{\prime} \}
	\ .
	\end{cases} 
\end{equation}
Therefore, to first order in the external field the time dependence of
the matrix elements of Eq. (\ref{q_34}) is:
\begin{equation}
\label{q_39}
	\begin{split}
	&
	\langle n^{\prime} \ell^{\prime} m^{\prime} | 
	{e}^{  i {\hat H} t} {\hat O}
	{e}^{- i {\hat H} t} 
	| n \ell m \rangle ^{(F)}
	\\
	& 
	= 
	F_{1}
	\!\!\!
	\sum_{ \scriptstyle{ \lambda \neq n^{\prime} } \atop \scriptstyle{ \lambda \neq n } }
	\!\!\!
	\langle n^{\prime} \ell^{\prime} m^{\prime} | {\hat O} | \lambda \rangle
	\langle \lambda | {\hat x} | n \ell m \rangle 
\\
& \times
	\frac{ 
	{e}^{ - i ( E_{n} - E_{n^{\prime}} ) t } - 
	{e}^{ - i ( E_{\lambda} - E_{n^{\prime}} ) t } 
	}
	{ E_{n} - E_{\lambda} } 
	\\
	&
	+ 
	F_{2}
	\!\!\!
	\sum_{ \scriptstyle{ \mu \neq n } \atop \scriptstyle{ \mu \neq n^{\prime} } }
	\!\!\!
	\langle n^{\prime} \ell^{\prime} m^{\prime} | {\hat x} | \mu \rangle
	\langle \mu | {\hat O} | n \ell m \rangle 
\\
& \times
	\frac{ 
	{e}^{ - i ( E_{n} - E_{n^{\prime}} ) t } - 
	{e}^{ - i ( E_{n} - E_{\mu} ) t }
	}
	{ E_{n^{\prime}} - E_{\mu} }
	\\
	&
	- i
	F_{3}
	\!\!\!
	\sum_{ \ell_{\lambda} , m_{\lambda} }
	\!\!\!
	\langle n^{\prime} \ell^{\prime} m^{\prime} 
	| {\hat O} | n \ell_{\lambda} m_{\lambda} \rangle
	\langle n \ell_{\lambda} m_{\lambda} | {\hat x} | n \ell m \rangle 
\\
& \times
	{e}^{ - i ( E_{n} - E_{n^{\prime}} ) t } t 
	\\
	&
	+
	F_{4}
	\!\!\!
	\sum_{ \ell_{\lambda} , m_{\lambda} }
	\! \!\!
	\langle n^{\prime} \ell^{\prime} m^{\prime} | {\hat O} 
	| n^{\prime} \ell_{\lambda} m_{\lambda} \rangle
	\langle n^{\prime} \ell_{\lambda} m_{\lambda} | {\hat x} | n \ell m \rangle 
\\
& \times
	\frac{ {e}^{ - i ( E_{n} - E_{n^{\prime}} ) t } - 1 }
	{ E_{n} - E_{n^{\prime}} }  
	\\
	&
	+ i
	F_{5}
	\!\!\!
	\sum_{ \ell_{\mu} , m_{\mu} }
	\!\!\!
	\langle n^{\prime} \ell^{\prime} m^{\prime} | {\hat x} 
	| n^{\prime} \ell_{\mu} m_{\mu} \rangle
	\langle n^{\prime} \ell_{\mu} m_{\mu} | {\hat O} | n \ell m \rangle 
\\
& \times
	{e}^{ - i ( E_{n} - E_{n^{\prime}} ) t } t
	\\
	&
	+ 
	F_{6}
	\!\!\!
	\sum_{ \ell_{\mu} , m_{\mu} }
	\!\!\!
	\langle n^{\prime} \ell^{\prime} m^{\prime} | {\hat x} | n \ell_{\mu} m_{\mu} \rangle
	\langle n \ell_{\mu} m_{\mu} | {\hat O} | n \ell m \rangle 
\\
& \times
	\frac{ {e}^{ - i ( E_{n} - E_{n^{\prime}} ) t } - 1 }
	{ E_{n^{\prime}} - E_{n} } 
	\\
	& 
	+ 
	\langle n^{\prime} \ell^{\prime} m^{\prime} | {\hat O} | n \ell m \rangle 
	{e} ^{ - i ( E_{n} - E_{n^{\prime}} ) t } \; ,
	\end{split}
\end{equation}
where we have attached subscripts to the field strength $F$ only for bookkeeping
purposes, and so $F_{1} = F_{2} = ... = F_{6} = F$.

Clearly, all the terms of Eq. (\ref{q_39}) oscillate with a frequency 
comparable (but not identical!)
to the Kepler frequency $\omega_{K}$ of the hydrogenic manifold 
at the center of the distribution of principal quantum numbers. We indicate
the principal quantum number of this special hydrogenic eigenmanifold as ${\bar n}$.
For very weak external fields,
the Kepler frequency is much larger than the Stark frequency of the motion.
This is true for the classical conditions on the fields of Eq. (\ref{q_5}) and also,
albeit in a much stronger form, for the usual quantum condition of Eq. (\ref{q_5c}),
i.e., the Inglis-Teller limit 
for negligible intermanifold mixing. 
This means that the wave packet oscillates several times before 
the classical perturbative equations of motion yield any significant
change in the expectation values of angular momentum and Runge-Lenz vector.
Therefore, following exactly the approach
of {\em classical} perturbation theory for the 
derivation of Eqs. (\ref{q_3}), we time average the 
quantum dynamics over the Kepler
period $T_{K} = 2 \pi {\bar n}^{3}$ of the ${\bar n}$-manifold 
approximately at the center of the 
energy distribution of the wave packet. 
Such time averaging does not affect the diagonal terms of Eq (\ref{q_30}); however,
it allows us to
evaluate with accuracy the {\em secular}, off-diagonal contributions to the 
dynamics over a time $\tau = \gamma T_{S}$, that is, over some multiple of the 
Stark period $T_{S}$.

For a Rydberg electron confined within a $n$-manifold
the Stark period $T_{S}$ is defined as:
\begin{equation}
\label{q_42aa}
	 T_{S} = \frac{ 2 }{ 3nF }  \; .
\end{equation}
However, its definition can be easily generalized
to the case of a wave packet with average principal quantum
number ${\bar n}$, by setting:
\begin{equation}
\label{q_42a}
	T_{S} \approx \frac{2}{3 {\bar n} F} \; .
\end{equation}

First, we consider the last term of Eq. (\ref{q_39}) which is of zeroth
order in the field; by inserting it in the 
perturbative equations of 
motion, i.e., in Eq. (\ref{q_30}), one obtains an
off-diagonal term which is of first order in the field.
By first time averaging over a Kepler period, 
and then integrating over a time $\tau$, and finally
using Eq. (\ref{q_42a} one obtains:
\begin{equation}
\label{43}
	\begin{split}
	\int_{0}^{\tau}
	F 
	&
	\langle n^{\prime} | {\hat O} | n \rangle 
	\left\langle
	{e} ^{ - i ( E_{n} - E_{n^{\prime}} ) t }
	\right\rangle _{K}
	d \tau^{\prime} \\
	& \approx
	\gamma \frac{2}{ 3 {\bar n} } 
	\langle n^{\prime} | {\hat O} | n \rangle 
	\frac{ 3 }{ {\bar n} }
	\left( \Delta_{{\bar n},n} - \frac{1}{2} \Delta_{n^{\prime},n} \right) \; ,
	\end{split}
\end{equation}
where $\big{\langle} ... \big{\rangle}_{K}$ indicates the time averaging over the 
Kepler period, and where we used the following result (derived in Appendix B):
\begin{equation}
\label{q_40}
	\begin{split}
	&
	\left\langle
	{e} ^{ - i ( E_{n} - E_{n^{\prime}} ) t }
	\right\rangle _{K} = 
	\frac{1}{T_{K}}
	\int_{0}^{T_{K}} 
	\!\!
	{e} ^{ \mp i ( E_{\imath} - E_{\jmath} ) t } dt 
	\\
	& =
	\frac{3}{{\bar n}} \left( \Delta_{{\bar n},\jmath} - \frac{1}{2} 
	\Delta_{\imath ,\jmath} \right) +
	O \left( \frac{ \Delta^{2} }{ {\bar n}^{2} } \right) \; .
	\end{split}
\end{equation}
From Eq. (\ref{q_32}) one has:
\begin{equation}
\label{q_44}
	| \langle n^{\prime} | {\hat O} | n \rangle |
	= 
	| \langle n^{\prime} | {\hat r}_{\imath} | n \rangle |
	\lesssim \frac{ {\bar n}^{2} }{ 2 | \Delta_{ n^{\prime} , n } | } \; ,
\end{equation}
which holds only to the leading order in ${\bar n}$, and where
a similar result can be obtained also in the case
${\hat O} = n {\hat p}_{\imath} {\hat r}_{\jmath}$
The final result is:
\begin{equation}
\label{q_45}
	\begin{split}
	\left|
	\int_{0}^{\tau}
	\right.
	&
	F 
	\left.
	\langle n^{\prime} | {\hat O} | n \rangle 
	\left\langle
	{e} ^{ - i ( E_{n} - E_{n^{\prime}} ) t }
	\right\rangle _{K}
	d \tau^{\prime} \right|
	\\
	& \lesssim
	\gamma
	\left|
	\frac{ 2 \Delta_{{\bar n},n} - \Delta_{ n^{\prime} , n } }
	{ 2 | \Delta_{ n^{\prime} , n } | }
	\right|
	\ll {\bar n} \; ,
	\end{split}
\end{equation}
where the last inequality follows under the very important assumption
which we introduced before, i.e., that the 
variance of the distribution 
of the Rydberg wave packet over the principal quantum numbers 
is much smaller than the average principal quantum number
of the wave packet.
We conclude from Eq. (\ref{q_45}) that first order,
off-diagonal contributions are very small when compared 
to the dominating principal quantum number of the wave packet; 
indeed, this is a sufficient condition
to neglect them completely,
because the quantum expectation values of angular
momentum and scaled Runge-Lenz vector range precisely from $-{\bar n}$ to ${\bar n}$.
Most importantly, however, one must also require that $\gamma \sim 1$,
i.e., the off-diagonal contribution remains small only up
to times comparable to the Stark period; for longer times
the secular effects build up and off-diagonal terms become
relevant.

Clearly, the analysis above does not yet yield any information about the 
precise condition that the external field $F$ must satisfy, so that
all off-diagonal terms remain negligible. 
To learn more about it, one must
analyze the off-diagonal contributions which are of second order in the 
field. We begin considering the terms of Eq. (\ref{q_39}) which are 
proportional to $F_{1}$. 
By inserting any of them in the equations of motion
for the expectation values of
angular momentum and Runge-Lenz vector, 
one obtains contributions which are
of second order in 
the external field; that is, to the leading order in ${\bar n}$ one obtains:
\begin{equation}
\label{q_45a}
	\begin{split}
	\left|
%%%%%%%%%%%%%%%%%%%%%%%%%%%%%%%%%%%%%%%%%%%%%%%%%%%%%%%%%%%%%%%%	
\vphantom{	\frac{ 
		\left\langle
		{e} ^{ - i ( E_{n} - E_{ n^{\prime} } ) t }
		\right\rangle _{K} 
		}
		{ E_{n} - E_{\lambda} }		}
%%%%%%%%%%%%%%%%%%%%%%%%%%%%%%%%%%%%%%%%%%%%%%%%%%%%%%%%%%%%%%%%	
	\int_{0}^{\tau}
	\!\!\!
	F^{2}
	\!\!
	\right.
	&
	\left.
	\langle n^{\prime} \ell^{\prime} m^{\prime} | {\hat O} | \lambda \rangle 
	\langle \lambda | {\hat x} | n \ell m \rangle 
	\frac{ 
	\left\langle
	{e} ^{ - i ( E_{n} - E_{ n^{\prime} } ) t }
	\right\rangle _{K} 
	}
	{ E_{n} - E_{\lambda} }
	d \tau^{\prime} \right| \\
	& \lesssim
	\gamma \left|
	\frac{ \Delta_{ {\bar n},{\langle n \rangle} } }
	{ 2 \Delta_{ n^{\prime},\lambda } \Delta^{2}_{ n,\lambda} } \right|
	F {\bar n}^{5} \, ,
	\end{split}
\end{equation}
where ${\langle n \rangle} = ( n + n^{\prime} ) / 2 $, and we have used the results of 
Eqs. (\ref{q_42a}), (\ref{q_40}) and (\ref{q_44}); we have also used the 
following result (see Appendix B):
\begin{equation}
\label{q_45b}
	\begin{split}
	\frac{1}
	{ E_{\imath} - E_{\jmath} } = 
	\frac{ {\bar n}^{3} }{ \Delta_{\imath,\jmath} }
	&
	\left\{
	1 + 
	\frac{3}{{\bar n}} 
	\left( \Delta_{\jmath,{\bar n}} + \frac{1}{2} \Delta_{\imath,\jmath} \right) 
	\right. \\
	&
	\left.
	+ O \left( \frac{ \Delta^{2} }{ {\bar n}^{2} } \right) 
	\right\} \; .
	\end{split}
\end{equation}
From Eq. (\ref{q_45a}) is finally possible to extract a necessary and
sufficient condition on the 
field strength. By requiring that the result of Eq. (\ref{q_45a}) is much
smaller than ${\bar n}$ one obtains:
\begin{equation}
\label{q_44a}
	F {\bar n}^{4} \ll 1 \; ,
\end{equation}
which is essentially the same as the classical condition of Eq. (\ref{q_5}),
as we had claimed before.

Clearly, the same analysis applies to the terms of Eq. (\ref{q_39})
which are proportional to $F_{2}$, and also to the oscillating part 
of the $F_{4}$ and $F_{6}$ terms. 

Therefore, we next turn our attention to the remaining terms
from Eq. (\ref{q_39}), that is, the contributions which are
proportional to $F_{3}$ and $F_{5}$
and also the non-oscillating parts of the $F_{4}$ and $F_{6}$ terms
as well. 
First, one needs the time averages of the time dependent factors in the
$F_{3}$ and $F_{5}$ sums, which are given by the following 
equation (see Appendix B):
\begin{equation}
\label{q_41}
	\begin{split}
	&
	\left\langle
	i t \ {e} ^{ - i ( E_{\imath} - E_{\jmath} ) t }
	\right\rangle _{K} =
	\frac{ i }{T_{K}}
	\int_{0}^{T_{K}} 
	\!\!\!
	t \ {e} ^{ - i ( E_{\imath} - E_{\jmath} ) t } dt 
	\\
	& 
	=
	- \frac{ {\bar n}^{3} }{ \Delta_{\imath,\jmath} }
	\left\{
	1 - 
	\frac{6}{{\bar n}} 
	\left( \Delta_{{\bar n},\jmath} - \frac{1}{2} \Delta_{\imath,\jmath} \right)
	\left( 1 + i \pi \Delta_{\imath,\jmath} \right) \right. \\
	&
	\left.
	+ O \left( \frac{ \Delta^{3} }{ {\bar n}^{2} } \right)  
	\right\} \; .
	\end{split}
\end{equation}
By inserting the 
time averaged $F_{3}$ and $F_{5}$ terms, along with the 
non oscillating parts of the $F_{4}$ and $F_{6}$ terms
in the equations of motion and
integrating over time one obtains 
(to the leading order in ${\bar n}$), an additional intermanifold
correction, which we denote as $G ( n^{\prime} \ell^{\prime} m^{\prime}; n \ell m )$:
\begin{equation}
\label{q_41a}
	\begin{split}
	&
	G ( n^{\prime} \ell^{\prime} m^{\prime}; n \ell m ) =
	\gamma T_{S}
	F
	\frac{ {\bar n}^{3} }{ \Delta_{ n,n^{\prime} } } \\
	&
	\times
	\left\{ F_{3}
	\!
	\sum_{ \ell_{\lambda} , m_{\lambda} }
	\!
	\langle n^{\prime} \ell^{\prime} m^{\prime} 
	| {\hat O} | n \ell_{\lambda} m_{\lambda} \rangle
	\langle n \ell_{\lambda} m_{\lambda} | {\hat x} | n \ell m \rangle
	\right.
	\\
	&
	- F_{4}
	\!
	\sum_{ \ell_{\lambda} , m_{\lambda} }
	\!
	\langle n^{\prime} \ell^{\prime} m^{\prime} 
	| {\hat O} | n^{\prime} \ell_{\lambda} m_{\lambda} \rangle
	\langle n^{\prime} \ell_{\lambda} m_{\lambda} | {\hat x} | n \ell m \rangle
	\\
	&
	- F_{5}
	\!
	\sum_{ \ell_{\mu} , m_{\mu} }
	\!
	\langle n^{\prime} \ell^{\prime} m^{\prime} | {\hat x} 
	| n^{\prime} \ell_{\mu} m_{\mu} \rangle
	\langle n^{\prime} \ell_{\mu} m_{\mu} | {\hat O} | n \ell m \rangle
	\\
	&
	\left.
	+ F_{6}
	\!
	\sum_{ \ell_{\mu} , m_{\mu} }
	\!
	\langle n^{\prime} \ell^{\prime} m^{\prime} | {\hat x} 
	| n \ell_{\mu} m_{\mu} \rangle
	\langle n \ell_{\mu} m_{\mu} | {\hat O} | n \ell m \rangle 
	\right\} \; .
	\end{split}
\end{equation}
It is easy to see from the previous analysis of the $F_{1}$ terms
that $G  ( n^{\prime} \ell^{\prime} m^{\prime}; n \ell m )   $ 
is a negligible intermanifold contribution if and only if the 
expression within the curly brackets of Eq. (\ref{q_41}) scales
as $\sim F {\bar n}^{3}$. However, each of the terms in the four sums consists
of the product of two matrix elements, and our previous analysis
of the magnitude of the matrix elements of ${\hat r}_{\imath}$ and also of any of
the possible choices for ${\hat O}$, indicates that all such terms
may scale as $\sim F {\bar n}^{4}$. Therefore, the desired scaling
as the third power of ${\bar n}$
must originate from cross cancelations between the
four sums within the curly brackets.

This is obviously correct when, for example, ${\hat O} = n {\hat L}_{z}$
and all the off-diagonal matrix elements of 
$G  ( n^{\prime} \ell^{\prime} m^{\prime}; n \ell m ) $ vanish, and one has:
\begin{equation}
\label{q_41b}
	\begin{split}
	& | G ( n^{\prime} \ell^{\prime} m^{\prime}; n \ell m ) | 
	\\
	& = 
	\gamma \frac{ 2 {\bar n}^{2} }{ 3 | \Delta_{ n,n^{\prime} } | }
	| F n \ ( m - m^{\prime} ) 
	\langle n^{\prime} \ell^{\prime} m^{\prime} | {\hat x} |
	n \ell m \rangle | 
	\\
	&
	\approx 
	\gamma \frac{ F {\bar n}^{5} }{ 3 \Delta_{ n,n^{\prime} } ^{2} }
	\end{split}
\end{equation}
where we used Eq. (\ref{q_42a}), and also
the usual selection rules to conclude that 
$m - m^{\prime} = \pm 1$. A moment's thought shows that an essentially
similar analysis holds also for the other components of the angular momentum.

The situation, instead, becomes much more complicated when ${\hat O} = {\hat r}_{\imath}$
or ${\hat O} = n {\hat p}_{\imath} {\hat r}_{\jmath}$ In those cases
the pairing of terms which leads
to the desired cross cancelations depends on the differences of the angular
momentum quantum numbers; that is, it depends on $\Delta_{\ell^{\prime} , \ell}$ and 
possibly also on 
$\Delta_{m^{\prime} , m}$. 
For example, if $\ell = \ell^{\prime} - 2$, by pairing an
$F_{3}$ term with the corresponding $F_{4}$ term one obtains: 
\begin{equation}
\label{q_41c}
	\begin{split}
	&
	\sum_{ \ell_{\lambda} , m_{\lambda} }
	\eta 
	\left( F_{3} R_{n^{\prime}, \ell^{\prime}}^{n, \ell_{\lambda}}
	R_{n, \ell_{\lambda}}^{n ,\ell} -
	F_{4} R_{n^{\prime}, \ell^{\prime}}^{n^{\prime}, \ell_{\lambda}}
	R_{n^{\prime}, \ell_{\lambda}}^{n ,\ell}  
	\right) \\
	& \approx
	\sum_{ \ell_{\lambda} , m_{\lambda} }
	\eta F 
	R_{n^{\prime}, \ell^{\prime}}^{n, \ell_{\lambda}}	
	\left( R_{n, \ell_{\lambda}}^{n ,\ell}  - 
	R_{n^{\prime}, \ell^{\prime}}^{n^{\prime}, \ell_{\lambda}}  
	\right) \; ,
	\end{split}
\end{equation}
where $\eta$ is a coefficient $\sim 1$ which contains the angular part
of the matrix elements {\cite{bethe_77}}.
The difference of the two diagonal radial matrix elements scales
as $\sim n \approx {\bar n}$, and therefore the whole expression scales as 
$\sim F {\bar n}^{3}$, which is the desired result. Note that 
in Eq. (\ref{q_41c}) one can approximately
factor out an off-diagonal radial matrix element 
because both off-diagonal elements represent the same kind of transition,
i.e., the principal quantum number {\em and} the angular momentum
quantum number decrease in both cases (assuming that $n^{\prime} > n$), that is
the transitions are:
\begin{equation}
\label{q_41ca}
	n^{\prime} \rightarrow n \; , \;
	\begin{cases}
	\ell^{\prime} \rightarrow \ell_{\lambda} &
	\ell^{\prime} > \ell_{\lambda} \\
	\ell_{\lambda} \rightarrow \ell &
	\ell_{\lambda} > \ell \; .
	\end{cases}
\end{equation}
Indeed, it is well known that the matrix element for an atomic transition
which increases both the energy and the angular momentum of the 
electron is significantly 
larger than the one for a transition which brings about 
the same change in energy, but leads to a smaller final angular momentum {\cite{bethe_77}}.
Therefore, if $\Delta_{\ell^{\prime}, \ell} = 0$ a different, more complicated
pairing of the terms
must be employed, which in this case may depend also on the angular part of the 
matrix elements.

However, the scaling of $G  ( n^{\prime} \ell^{\prime} m^{\prime}; n \ell m ) $ 
with the principal quantum number can most 
effectively and also more convincingly be studied by evaluating numerically the 
whole expression within the curly brackets of Eq. (\ref{q_41a}), divided by
F. More precisely, for each pair $\{ n^{\prime} , n \}$ we computed the 
maximum magnitude of the expression within curly brackets (divided by F)
over all possible choices of angular quantum numbers, and we 
we denote it by
$g( n^{\prime}, n )$, that is:
\begin{equation}
\label{q_41d}
	g( n^{\prime}, n ) = 
	{ \underset{ \{ \ell^{\prime} m^{\prime} ; \ell m \} }{\text max} } 
	\left|
	\frac{ \Delta_{n, n^{\prime}} 
	G ( n^{\prime} \ell^{\prime} m^{\prime}; n \ell m ) }
	{  \gamma T_{S} F^{2} {\bar n}^{3} }
	\right| \; .
\end{equation}
%
%%%%%%%%%%%%%%%%%%%%%%%%%%%%%%%%%%%%%%%%%%%%%%%%%%%%%%%%%%%%%%%%%%%%%%%%%%%%%%%
%
%%%%\begin{comment}
%
% Importing a Figure
%
\begin{figure}
\centerline{\psfig{file=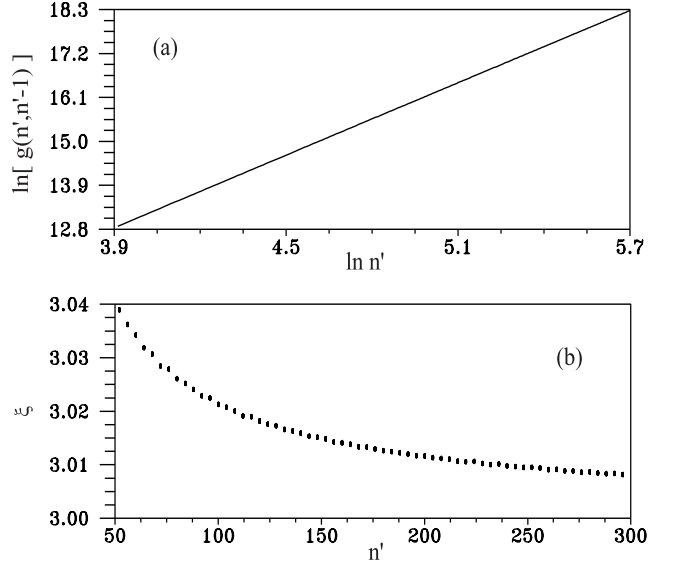,height=7.5cm,width=8.5cm,angle=0}}
\caption{
Scaling of the intermanifold contribution with the principal
quantum number. 
In Fig (a) we plot the natural logarithm of the maximum magnitude 
of the intermanifold terms vs. the natural logarithm
of the principal quantum number, and the approximate straight 
line indicates a simple power-law scaling.
In Fig (b) we plot the slope of the line in Fig. (a), i.e.,
the exponent of the power-law, which is clearly converging to 
$\xi = 3$, thereby proving that in the semiclassical limit all
intermanifold contributions become negligible.
}
\label{q_fig_1}
\end{figure}
%
%%%%\end{comment}
%
%%%%%%%%%%%%%%%%%%%%%%%%%%%%%%%%%%%%%%%%%%%%%%%%%%%%%%%%%%%%%%%%%%%%%%%%%%%%%%%
%
In Fig. \ref{q_fig_1}-(a) we plot the logarithm of $g ( n^{\prime} , n^{\prime} - 1 )$
vs. the logarithm of $n^{\prime}$, setting ${\hat O} = {\hat x}$:
this is the case in which 
$ G  ( n^{\prime} \ell^{\prime} m^{\prime}; n \ell m ) $ is the largest, as 
many numerical calculations confirm.
The line in Fig. \ref{q_fig_1}-(a) is almost exactly a straight,
which confirms that 
$ g  ( n^{\prime} , n^{\prime}  - 1) $ scales with 
$n^{\prime} \approx {\bar n}$ according to a 
power law, i.e. $ g \sim {\bar n}^{\xi} $. 
However, the exponent $\xi$ is not exactly constant, and 
in Fig. \ref{q_fig_1}-(b) we plot
the slope of the straight line of Fig. \ref{q_fig_1}-(a) vs. $n^{\prime}$.
Clearly, for increasing $n^{\prime}$'s the exponent ${\xi}$
is converging to 3, and that is precisely the result which we need
to prove that 
$ G  ( n^{\prime} \ell^{\prime} m^{\prime}; n \ell m ) $  
is truly a negligible intermanifold contribution to the equations of motion.

The proportionality coefficient of the power law
can also be easily evaluated from the numerical
data and one obtains:
\begin{equation}
\label{q_41e}
	g( n^{\prime}, n^{\prime} - 1 ) \approx 3.15 {n^{\prime}}^{3}
	= 3.15 {\bar n}^{3} \left\{ 1 + O \left( \frac{ \Delta }{ {\bar n} } \right) 
	\right\} \; .
\end{equation}
Finally, we have repeated the same calculations for several values
of $\Delta_{n^{\prime}, n}$, and also for ${\hat O} = {\hat y}$, ${\hat z}$,
and ${\hat O} = n {\hat p}_{\imath} {\hat r}_{\jmath}$, $\{\imath,\jmath\} = 1,2,3$.
In all cases our findings were essentially identical to the ones of Fig. \ref{q_fig_1}.

The physical interpretation of our result is particularly interesting.
Rydberg wave packets which are relatively well localized in energy
move along the trajectories of the classical 
electron or, for radial wave packets, of an ensemble of classical electrons
{\cite{%
carlos_86a,carlos_91a,carlos_94a,carlos_94b,carlos_94c,%
carlos_95a,carlos_96a,carlos_97a,%
perelman_89,%
nauenberg_89a,nauenberg_90a,nauenberg_94a,%
eberly_94a,eberly_95a,eberly_96a,eberly_96b,%
turgay_95a,turgay_96a,turgay_97a,david_97a,turgay_98a,%
bialynicki_97a,delande_95a,delande_95b}.
However, unless some suitable external fields are applied to the system,
{\cite{%
eberly_94a,eberly_95a,eberly_96a,eberly_96b,%
turgay_95a,turgay_96a,turgay_97a,david_97a,turgay_98a,%
bialynicki_97a,delande_95a,delande_95b},
they do so only for a few Kepler periods.
Such wave packets do not remain 
localized in the angular variable and therefore
spread along the classical ellipse. 
Eventually they display interference 
fringes as the front of the packet catches up with its tail, and finally they also 
show quantum revivals and superrevivals 
{\cite{carlos_86a,carlos_91a,carlos_94a,carlos_95a,carlos_97a,perelman_89,nauenberg_89a}}.
However, from the point of view of the time averaged equations of motion the 
electron is {\em always} spread (i.e., averaged) 
along the classical trajectory, very much like in
classical mechanics, where after time averaging
the elements of the Kepler ellipse become 
the dynamical variables of the 
system, replacing the phase space coordinates of the electron.
Therefore the time averaged, quantum equations of motion are insensitive to
the spreading of the wave packet, and to its revivals, and that is why the 
same results as for stationary states can be legitimately
extended to wave packets too.

Note that the results of our analysis can be generalized beyond the case of 
Rydberg electrons excited to wave packets, and hold also when
the Rydberg electron is initially confined within a hydrogenic manifold,
and $n$-mixing is brought about only by external field. In that 
case, in fact, one needs only
to replace $\langle n^{\prime} \ell^{\prime} m^{\prime} |$ with 
$\langle n \ell^{\prime} m^{\prime} |$.
Clearly, for a Rydberg electron initially confined within a
$n$ manifold, all off-diagonal terms in the 
equations of motion for the quantum expectation values
are of second order in the field, however they still derive from
the first order terms of Eq. (\ref{q_39}), and that is why the
classical constraint on the external field of Eq. (\ref{q_44a}) 
holds in that case too.

Indeed, in the very important case of slow ion-Rydberg collisions
the Rydberg electron is initially excited to a specific $n$-manifold, 
and in the next section we show that some special linear combinations
of hydrogen atom eigenfunctions which are confined within a hydrogenic manifold
(elliptic states {\cite{delande_89a,delande_90a,nauenberg_89a}}) 
yield the appropriate quasiclassical initial conditions for 
the quantum expectation values of angular momentum and Runge-Lenz vector,
which then closely track the time averages of the classical variables.

\section{classical, quasiclassical and quantum initial conditions}

In this section we discuss the initial conditions for the 
equations of motion for the quantum expectation values,
and show that 
in the case of elliptic states 
{\cite{delande_89a,delande_90a,nauenberg_89a}} the
quantum expectation values track exactly the time averages 
of the classical variables. 
More precisely, for a classical Kepler ellipse, initially in the {\em xy}-plane, and with
the semimajor axis pointing along the {\em x}-axis, one has:
\begin{eqnarray}
\label{q_50}
	L_z & = n \sqrt{ 1 - e^2 } \; , \; a_x = ne \nonumber \\
	L_x & = L_y = a_y = a_z = 0 \; & , \; 
\end{eqnarray}
where $e$ is the eccentricity of the orbit, and, as usual, the energy of the 
classical electron is: 
\begin{equation}
\label{q_51}
	E = - \frac{1}{2 n^{2} } \; .
\end{equation}
On the other hand, for an 
elliptic state $|n\alpha \rangle$ (an elliptic state is given by a 
complicated superposition of spherical eigenstates of the hydrogen atom, all
with the same principal quantum number
{\cite{delande_89a,delande_90a,nauenberg_89a}}), 
which is also localized in the {\em xy}-plane,
and oriented like the classical ellipse above, 
one has {\cite{delande_89a,delande_90a,nauenberg_89a}}:
\begin{equation}
\label{q_52}
	\begin{split}
	& \langle n\alpha | {\hat L}_z | n\alpha \rangle = (n-1) \cos \alpha \\
	& \langle n\alpha | {\hat a}_x | n\alpha \rangle = (n-1) \sin \alpha \\
	& \langle n\alpha | {\hat L}_x | n\alpha \rangle =  
	  \langle n\alpha | {\hat L}_y | n\alpha \rangle = 0 \\
	& \langle n\alpha | {\hat a}_y | n\alpha \rangle =   
	  \langle n\alpha | {\hat a}_z | n\alpha \rangle = 0 	\; ,
	\end{split}
\end{equation}
so that the correspondence between $e$ and $\sin \alpha$ is established 
(obviously this $\alpha$ has no relation with the coefficients of the previous section).
Clearly, in the limit of large {\em n}'s the quantum expectation values and
the classical predictions converge,
and for elliptic states the quantum expectation values not only obey the same
perturbative equations as the time averages of the classical variables,
but also have almost the same initial conditions. Therefore
they closely follow the same trajectories as the 
time averaged classical angular momentum and Runge-Lenz vector.
This result has already been 
observed numerically {\cite{gross_86a,carlos_97c,carlos_98a}}, and also
experimentally {\cite{gross_86a,horsdal_98a}} for some
special configurations of the external fields.
Most importantly, since elliptic states are 
coherent states of the angular momentum
{\cite{perelomov_86,delande_89a,delande_90a,nauenberg_89a}},
i.e., states of minimum uncertainty, it turns out that:
\begin{equation}
\label{q_53}
	\begin{split}
		& \frac{ 
		  \langle n\alpha | {\hat L}^{2} | n\alpha \rangle -
		  \sum_{\imath} \! \langle n\alpha | {\hat L}_{\imath} | n\alpha \rangle ^{2} 
  		  }{
    		  \frac{1}{2} \left( 
		  \langle n\alpha | {\hat L}^{2} | n\alpha \rangle +
		  \sum_{\imath} \! \langle n\alpha | {\hat L}_{\imath} 
		  | n\alpha \rangle ^{2} \right) 
		  } \\
		& = \frac{1}{ ( n - 1 ) \cos^{2} \alpha + \frac{1}{2} }	\; .
	\end{split}
\end{equation}
and in the semiclassical limit the expectation values of ${\hat L}_{\imath}$, $\imath=1,2,3$,
are related to the expectation value of ${\hat L}^{2}$ 
approximately like in classical mechanics.
In fact, it has been verified numerically in a few 
special cases {\cite{carlos_97c,carlos_98a}} that not only the 
expectation values of ${\hat L}_{\imath}$ and ${\hat a}_{\imath}$ evolve
in time quasiclassically
according to Eqs. (\ref{q_3}), but also that  during the time evolution
the state remains elliptic.
This is exactly the same situation as 
in classical mechanics, where
the electron keeps moving along an ellipse, but the properties of 
the ellipse vary slowly in time; similarly, the numerical evidence
shows that the elliptic state remains localized along a
classical Kepler ellipse while it slowly evolves in time.

Instead, for the more familiar spherical eigenstates $|n \ell m \rangle $'s
of the hydrogen atom the situation is completely different, 
and one has:
\begin{equation}
\label{q_54}
	\begin{split}
	& \langle n\ell m | {\hat L}_z | n\ell m \rangle = m \\
	& \langle n\ell m | {\hat L}_x | n\ell m \rangle = 
	  \langle n\ell m | {\hat L}_y | n\ell m \rangle = 0 \\
	& \langle n\ell m | {\hat a}_x | n\ell m \rangle = 
	  \langle n\ell m | {\hat a}_y | n\ell m \rangle = \\
	& = \langle n\ell m | {\hat a}_z | n\ell m \rangle = 0  \; ,
	\end{split}
\end{equation}
and the initial conditions differ dramatically from the classical ones, which leads
to some interesting considerations.

As we explained before, the classical constraints of Eq. (\ref{q_5a}) remain
invariant under the evolution of the perturbative equations. The second 
classical constraint
translates into a condition over the quantum expectation values,
which is also invariant:
\begin{equation}
\label{q_55}
	{\chi} = \sum_{\imath} \left\{  \langle \psi | {\hat L}_{\imath} | \psi \rangle^{2} +
	\langle \psi | {\hat a}_{\imath} | \psi \rangle^{2} \right\}	\; ,
\end{equation}
where $| \psi \rangle$ is 
any state (elliptic, spherical or also a wave packet, in which case 
the expectation value must also be averaged over a Kepler period)
which satisfies the requirements of our derivation. 
The invariant ${\chi}$ is related to the Casimir operator of the SO(4) symmetry group
of the hydrogen atom {\cite{englefield_72,wybourne_74}}.
The value of the classical invariant of Eq. (\ref{q_5a}) is $n^{2}$.
For an elliptic state the quantum invariant of Eq. (\ref{q_55})
is equal to $(n-1)^2$; however, for a spherical 
eigenstate $\chi$ is equal to $m^2$, which for the small $m$'s typically excited by 
optical transitions from the initial 
low-$n$ state to the Rydberg high-$n$ state
is a much smaller number than the classical result.
This poses severe limits on the largest possible expectation value of any component 
of the angular momentum over a spherical eigenstate;
this feature might be exploited in experiments to study the properties of
Rydberg states.

The limitation which Eq. (\ref{q_55}) imposes on 
the expectation values of angular momentum and Runge-Lenz vector
over spherical eigenstates 
stems from the fact that such states have vanishing electric dipole moment;
a small angular momentum is not balanced by a large Runge-Lenz vector,
as it happens in classical ellipses and quantum elliptic states.
More precisely, for spherical states, the 
electric field cannot induce first order dynamical effects 
because the expectation value of the Runge-Lenz vector over a spherical
eigenstate is zero (i.e., there is no permanent electric dipole), and 
the state must first be distorted by the field so 
that the expectation value of the angular momentum (or
Runge-Lenz vector) can change. This indicates that the dynamics must be at least of 
second order in the external fields.
This situation is germane to the well known
linear Stark effect {\cite{bethe_77}, where degenerate perturbation theory
and parabolic states must be used to account for the linear dependence
of the eigenvalues on the external field.
In fact, a spherical eigenstate can be seen as 
a superposition of elliptic states (or in a semiclassical 
interpretation an ensemble of Kepler ellipses), which are
oriented uniformly in the {\em xy}-plane,
so that the total Runge-Lenz vector is averaged to zero.

The (at least) quadratic dependence of the time evolution on the field 
can be explicitly verified by expanding the time
dependent operators in the Heisenberg picture,
{\cite{sakurai_85}} and by showing that the expectation values of the
first order terms in the electric field vanish.
Using the Hamiltonian of Eq. (\ref{q_1}) (minus the diamagnetic term),
and writing the hydrogen atom Hamiltonian as ${\hat H}_{0}$
one has:
\begin{equation}
\label{q_56}
	\begin{split}
	& e^{ i {\hat H} t } {\hat L}_{\imath} e^{ - i {\hat H} t } =
	{\hat L}_{\imath} - i t  
	\left[ {\hat H}_{0} - \omega_{L} {\hat L}_{z} , {\hat L}_{\imath} \right]
	- i t  F \left[ {\hat x} , {\hat L}_{\imath} \right] \\
	& + F \sum_p \! \frac{ ( - i t )^p }{ p ! } \! \sum_{k=0}^{p-1}
	\left[ {\hat H}_0 - \omega_L {\hat L}_z, \left[ ...
	\left[ x , \left[ {\hat H}_0 - \omega_L {\hat L}_z, 
	\left[ ... \right.\right.\right.\right.\right.     \\
	& \left.\left.\left.\left.\left.     ...
	\left[ {\hat H}_0 - \omega_L {\hat L}_z, {\hat L}_{\imath} \right] ...
	\right] \right] \right] \right] \right] +
	O ( F^2 ) \; ,
\end{split}
\end{equation}
The simplest first order term is a direct commutator of the field with a 
component of the angular momentum: therefore it is either zero or a component 
of the position operator (depending on the index ${\imath}$), 
and its expectation value over a spherical
eigenstate vanishes. 
Other, more complex first order terms come from 
the double sum in Eq. (\ref{q_56}) and consist of a
first series of $k$ commutators of ${\hat L}_{\imath}$
with ${\hat H}_{0} - \omega_L {\hat L}_{z}$, 
and then of a commutator with the field, 
and finally of
a second series of $(p-1-k)$ commutators
with ${\hat H}_{0} - \omega_L {\hat L}_{z}$.
It is easy to see that the expectation value of any of such terms over 
a spherical eigenstate 
of the hydrogen atom which is quantized along the {\em z}-axis vanishes, simply because
such state is an eigenstates of ${\hat H}_{0} - \omega_{L} {\hat L}_{z}$. 
However, the same result can be proven as follows also when the atom is quantized 
along an arbitrary direction.
I) The first series of $k$ commutators either vanishes or yields a component of the 
angular momentum (depending on the index ${\imath}$); 
II) the commutator of the result of step I
with the electric field either vanishes or yields a component of the position 
operator; III) finally,  the
result of steps I and II  must be commuted 
$(p-1-k)$ times with
with ${\hat H}_{0} - \omega_{L} {\hat L}_{z}$; this sequence of
commutators can be organized so that one does first the commutators with ${\hat L}_{z}$
and next those with ${\hat H}_{0}$
because these two operators commute
with one another. 
Obviously, the expectation value 
of the commutator of any operator with ${\hat H}_{0}$ vanishes, if 
it is taken over an 
eigenstate of ${\hat H}_{0}$ itself. 
On the other hand the commutators with ${\hat L}_z$ either vanish 
directly or yield
a component of the position operator, whose expectation value over a 
spherical state also vanishes, and our point is proved.

Similar considerations apply also to ${\hat L}^{2}$. In fact, it is even easier to 
demonstrate that the time evolution of the expectation value of ${\hat L}^{2}$
over a spherical eigenstate of the hydrogen atom is of second order in the field.
Clearly, $|n \ell m\rangle $ is an eigenstate of ${\hat L}^{2}$ and the
expectation value of the commutator of ${\hat x}$ with ${\hat L}^{2}$  over
$ | n \ell m \rangle $ vanishes. Moreover,
${\hat L}^{2}$ commutes with ${\hat H}_{0} - \omega_{L} {\hat L}_{z}$ and therefore
the sequence of commutators similar the one of Eq. (\ref{q_56}) can be rearranged so that
first one commutes the field operator with ${\hat H}_{0} - \omega_{L} {\hat L}_{z}$,
and next the result is commuted with ${\hat L}^{2}$. However, 
the expectation value over $| n \ell m\rangle $ of the
commutator of any operator with ${\hat L}^{2}$ vanishes, which proves our point. Clearly,
this result does not depend on the orientation of the axis of quantization
of the atom relative to the external fields.

Note that even if the time evolution of the expectation values of ${\hat L}_{\imath}$
and ${\hat L}^{2}$ is only of second order in the external field, that does not
imply that weak external fields are not effective in bringing about 
changes of the angular momentum. For example, in the expansion of Eq. (\ref{q_56})
the second order terms are multiplied by at least a square power of the time;
if we consider times comparable to the Stark period and the 
scaling of the matrix elements of the position operator with the principal
quantum number as given in Eq. (\ref{q_31a}), it is easy to see that
the final result is
not negligible. Moreover, the constraint on the 
quantum invariant $\chi$ of Eq. (\ref{q_55}) does not say much about
the total angular momentum of a spherical state, which is not a coherent state
of the angular momentum, and therefore not only one has:
\begin{equation}
\label{q_61}
	\sum_{\imath} \langle n \ell m | {\hat L}_{\imath} | n \ell m \rangle^{2} \neq
	\langle n \ell m | {\hat L}^{2} | n \ell m \rangle
\end{equation}
but the difference between the two sides of the equation can be very large, as
one can see by considering a state with $\ell = n -1$ and $m=0$.

Since the results of the previous section show that 
under the classical conditions for the external fields intermanifold
contributions to the dynamics can be neglected, all
our considerations apply also to a 
superposition of spherical states with different {\em n}'s, 
and therefore our analysis sheds some light on the
nature of the Rydberg states employed in ZEKE spectroscopy. 

In ZEKE, ultrahigh
molecular Rydberg states 
{\cite{vrakking_95a,vrakking_95b,vrakking_95c,reiser_88,zeke_0,zeke_1,zeke_5}} 
are first excited by a few
optical transitions and successively field ionized. 
This technique is extremely successful because of the ultralong 
lifetimes of these Rydberg states, which are explained in terms of 
extensive intrashell mixing of the initial, unstable low-$\ell$
states with the longer lived high-$\ell$ states. 
For increasing angular momenta
the coupling between the Rydberg electron and the molecular
core becomes rapidly negligible, so that autoionization and predissociation
channels are effectively quenched, and the Rydberg state becomes ultralong-lived. 
Therefore it is understood that ZEKE states, i.e., the ultralong-living
Rydberg states responsible for the ZEKE signal, are complicated superpositions
of large-$n$ spherical eigenstates of the hydrogen atom, which are skewed in favor of 
large angular momentum states.
Because of the small spacing of high-{\em n} Rydberg eigenenergies and of the 
width of the initial laser pulses, ZEKE states initially consist of a superposition
of several states with different principal 
quantum numbers {\cite{chupka_93a,chupka_93b}}. 
However, it is generally assumed that
only one angular momentum quantum number is allowed in the superposition because
of the usual selection rules. The population of higher-$\ell$ states is 
ascribed solely to the effect of external fields.

In fact, several
experimental studies 
{\cite{vrakking_95a,vrakking_95b,vrakking_95c,merkt_97a}}
have shown that the vanishingly small
stray fields of the experimental set-up
and, most importantly, the very weak, slowly
varying electric fields of the ions present in
the interaction region populate with great efficacy the
high-$\ell$ Rydberg states which are 
responsible for the observed ultralong lifetimes of ZEKE states.
On the theoretical side recent 
results {\cite{chupka_93a,chupka_93b,myself_6,myself_7,myself_8,myself_9}}, some of which
were based on the classical perturbative approach of Eqs. (\ref{q_3})
{\cite{myself_6,myself_7,myself_8,myself_9}},
have explained 
$\ell$-mixing in terms of the hydrogenic model, in which 
vanishingly small fields are sufficient to induce the desired 
scrambling of the angular momentum quantum numbers.
The great effectiveness with which such extremely weak fields
($F \lesssim 1 V / m$) populate high-$\ell$ states
strongly suggests that the hydrogenic model is indeed appropriate
to describe angular momentum mixing in ZEKE states. Moreover, our present findings
show that the previous classical results 
{\cite{myself_6,myself_7,myself_8,myself_9}} are really quantum mechanical in nature,
and can also be extended to the case of wave packets.
On the other hand, the low-$\ell$ states excited by the 
laser pulse have a nonnegligible quantum defect, which decouples
them from the high-$\ell$, quasi-hydrogenic states; 
it is then likely that another mechanism is at work. 
More precisely, it is possible that the initial optical
excitation of the ultrahigh-{\em n} states 
may not be strictly limited by the standard selection rules.
Instead, by contributions which are of higher order in the 
optical field and yet are nonnegligible because of the ultralarge dipole 
moments of Rydberg states -see Eq. (\ref{q_31a}) and Eq. (\ref{q_31b})-, the initial 
optical pulse may well populate a few angular
momentum states with relatively large $\ell$'s, 
as one of us has recently shown {\cite{carlos_97d}}.
Therefore, some degree of angular momentum mixing is probably already present in
the initial Rydberg state, in which case the hydrogenic model,
in its present extension to superpositions of states
with different quantum numbers, provides an accurate description
of how weak stray and ionic fields 
bring about the (approximate) randomization in $\ell$ and $m$ of ZEKE states,
which accounts for the observed
ultralong lifetimes.
 
Finally, the extension of equations Eq. (\ref{q_3}) from purely
classical variables to quantum expectation values lends strength to a
previous argument of ours concerning slow, ion-Rydberg collisions
and which until now was based solely on purely 
classical calculations {\cite{myself_7,myself_9}}.
More precisely, we suggested the need
for a review of both experimental and theoretical
results for the intrashell transitions induced in Rydberg alkali atoms
by slow collisions with ions.
In the case of slow ion-Rydberg collisions the ``magnetic" term 
of the Hamiltonian arises
from the rotation frequency of the field,
and the problem is treated in the 
frame rotating with the field itself.
In that frame the Hamiltonian is 
equivalent to the one of a hydrogenic electron 
in weak electric and magnetic fields of constant
orientation, and time dependent 
magnitude {\cite{demkov_74,myself_7,myself_8,myself_9}}. 
The ratio of the two magnitudes,
however, remains constant, and Eq. (\ref{q_3}) can be solved 
exactly {\cite{myself_7,myself_8,myself_9}}. 
It is then easy to see 
that after a full collision, the expectation value
of ${\hat L}_z$ is over a spherical eigenstate (that is with the initial conditions 
of Eq. (\ref{q_54})), is {\cite{myself_7,myself_9}:
\begin{equation}
\label{q_65}
	\lim_{t \rightarrow \infty} \langle n \ell m | {\hat L}_{z} (t) 
	| n \ell m \rangle =
	\frac{ 4 b^2 {\tilde v} - 9 n^2 }{4 b^2 {\tilde v} + 9 n^2 } m
\end{equation}
where {\em b} is the impact parameter of the collision, and ${\tilde v}$ is the 
``reduced" velocity of the incoming ion,  {\em i.e.} its velocity 
in atomic units multiplied
by {\em n}, which is the principal quantum 
number of the Rydberg electron in the target.
In a first approximation, one may insert in Eq. (\ref{q_65})
an average impact parameter $b \sim 50 n^2$
and a reduced velocity ${\tilde v} \sim 1 $, which are consistent with the 
experimental conditions, and the coefficient 
multiplying $m$ becomes $\gtrsim 0.9$.
Therefore the expectation value of ${\hat L}_z$ cannot change much, and if initially
the electron is prepared in a $m \neq 0$ state, the expectation value of ${\hat L}_z$
will not vanish.
However, both in the interpretation of the experimental data {\cite{macadam_93}}, 
and also in fully quantum theoretical 
treatments {\cite{syrkin_85,macadam_95,syrkin_96,cavagnero_95}}
the assumption has been made of a 
uniform population of the $m$ substates, which corresponds to 
a zero expectation value of ${\hat L}_z$.
For high $\ell$ states this is a reasonably good approximation, 
even in the case of a non vanishing 
expectation value of ${\hat L}_z$.   
However the approximation clearly
breaks down for smaller values of $\ell$, which is precisely the
regime for which we suggested a critical review of current results.

\section{conclusions}

In this paper we have shown 
that under realistic conditions, the classical and quantum dynamics of 
Rydberg electrons in weak, slowly varying external fields agree beyond the 
mandates of Ehrenfest theorem.

We have shown
that for the hydrogen atom in weak, slowly varying
electric and magnetic fields, to first order in the 
applied fields the quantum expectation values of the components of 
the angular momentum and the Runge-Lenz vector obey
exactly the same equations as the time averages (over a Kepler period
and along a Kepler ellipse)
of the corresponding classical variables.
Our proof follows in spirit the approach of classical perturbation theory, as
we fully exploit the 
properties of the zeroth order solutions of the quantum problem,
exactly as one does in classical mechanics where the time averaging is
done along Kepler ellipses, i.e., the zeroth order solutions
of the classical problem.
Note that 
this result is not an application of Ehrenfest's theorem, because
the perturbative approach does not consist of the linearization of the 
problem in the neighborhood of an equilibrium point. Instead, it is an
extension, in stronger form, of the theorem for the important case of 
the hydrogen atom in weak external fields.

Most importantly, in our derivation we have not applied Pauli's replacement
directly in the Hamiltonian
{\cite{nauenberg_94a,carlos_97c,carlos_98a}},
and therefore we have been
able to investigate the intermanifold contributions to the dynamics.
In fact, by time averaging the dynamics over a Kepler period (which, again, is 
exactly the same
procedure as in classical perturbation theory)
we have shown that 
intermanifold terms do not contribute
significantly to the evolution of the quantum expectation values, as long
as the strength of the external fields satisfies the same
requirement as in classical mechanics. Interestingly,
in the semiclassical limit the
classical constraint is much weaker than the quantum condition
for negligible $n$-mixing, i.e., the Inglis-Teller limit.

This paradox can be resolved by observing that the perturbative equations
remain accurate only up to times comparable to the 
Stark period, that is, for times which in atomic units are $\sim 1 / \Delta E$,
where $\Delta E$ is the energy separation
between two Stark levels. Therefore, over such
relatively short times the stationary picture of the Stark eigenstates 
which spread out
of a hydrogenic $n$-manifold does not have much physical significance,
and that is why for these relatively short times
the more stringent quantum condition can 
be ignored.

Moreover, by time averaging the intermanifold dynamics we have also extended the
validity of the classical perturbative equations to the
case of Rydberg wave packets, as long as the spread of the
packet over the hydrogenic eigenmanifolds is small
compared to its average principal quantum number.
Note, however, that although our analysis shows that
the quantum expectation values of angular momentum and Runge-Lenz vector
evolve in time like the classical time averaged variables, 
{\em it says nothing
about the localization of the wave packet
and the quasiclassical dynamics of the packet itself}.
In fact, our time averaging is precisely equivalent to 
considering a spread out version of the wave packet, 
smeared along its orbit. This is the same situation as in classical mechanics,
where one studies the motion of the Kepler ellipse, as if it
the classical electron had been magically smeared along its own 
trajectory.

We have also demonstrated that the close quantum-classical 
equivalence can be extended, in the limit
of very large principal quantum numbers, to the 
initial conditions of the equations of motion, provided that the 
expectation values are taken over elliptic states, which are 
states localized along the classical solutions 
{\cite{delande_89a,delande_90a,nauenberg_89a}}. 
Therefore the quantum expectation values of angular momentum
and Runge-Lenz vector over elliptic states follow essentially
the same trajectories as the time averages of the corresponding 
classical variables. Such complete quantum-classical equivalence,
however, does not hold for the more familiar spherical
eigenstates ($| n \ell m \rangle $) of the hydrogen atom.

The realization that the hydrogenic, perturbative equations of motion
(which account so well for several physical phenomena) can also be interpreted
as purely quantum mechanical equations has led to
some insight into the nature the Rydberg states employed in ZEKE spectroscopy; it
also lends support to our result (previously only classical) which indicates that the
averaging over the ${\hat L}_z$ sublevels (which is used in quantum close-coupling
calculations to the end of making the problem of ion-Rydberg collisions numerically
more tractable, and also in the interpretation of experimental data) 
may be unjustified.

Finally, one may wonder if the special equivalence between the dynamics of 
the time averages of classical variables and quantum expectation values is 
a peculiarity of the hydrogen atom in weak external fields, or if it can be 
extended to other weakly perturbed integrable systems, and the investigation
of this problem is in progress in our groups.

\appendix

\section{proof of the identity of Eq. (10)}

%%%%%%%%%%%%%%%%%%%%%%%%%%%%%%%%%%%%%%%%%%%%%%%%%%%%%%%%%%%%%%%%%%%%%%%%%%%%%%%%%%%%%%%%%%%%%%%
%
In our proof of the special quantum-classical equivalence of the 
dynamics of Rydberg electrons in weak external fields,  we have made extensive use
of the following identity:
\begin{equation}
\label{q_600}
	\langle \psi_{n} | {\hat r}_{\imath} {\hat p}_{\jmath} | \psi_{n} \rangle = 
	- \langle \psi_{n} | {\hat p}_{\imath} {\hat r}_{\jmath} | \psi_{n} \rangle \; ,
\end{equation}
where ${\hat r}_{\imath}$ and ${\hat p}_{\jmath}$ are 
components of the position and 
momentum operator respectively, and
where $| \psi_{n} \rangle$ is a state confined within a hydrogenic $n$-manifold. 

In this appendix we prove explicitly the identity of Eq. (\ref{q_600}),
and we do so
for all the pairs of indexes $\{ \imath, \jmath \}$
to stress that our derivation of the equations of motion
does not depend on the relative
orientation between the initial axis of quantization of the atom and the 
direction of the applied, external fields in the Hamiltonian of
Eq. (\ref{q_1}).

We begin with the simplest case, that is when $\imath=\jmath$:
\begin{equation}
\label{q_7}
	\begin{split}
	\langle \psi_{n} | 
	{\hat x_{\imath} } {\hat p_{\imath} } 
	&
	+ {\hat p_{\imath} } {\hat x_{\imath} }
	| \psi_{n} \rangle \\
	& 
	= - i \langle \psi_{n} | 
	{\hat x_{\imath} } \left[ {\hat x_{\imath} } , {\hat H_{0} } \right]
	+ \left[ {\hat x_{\imath} } , {\hat H_{0} } \right] {\hat x_{\imath} }
	| \psi_{n} \rangle \\
	&
	= - i \langle \psi_{n} | 
	\left[ {\hat x_{\imath} }^{2} , {\hat H_{0} } \right]
	| \psi_{n} \rangle = 0 \; ,
	\end{split}
\end{equation}
where ${\hat H_{0}}$ is the hydrogen atom Hamiltonian and the result follows
because $| \psi_{n} \rangle$ is an eigenstate of ${\hat H_{0}}$.
The same approach could be easily extended to all cases. 
However, for the cases in which $\imath \neq \jmath$ 
a different approach is more convenient to the end of 
studying the intermanifold contributions to the
equations of motion, which we do in the main text of the 
paper [see Eq. (\ref{q_31})]. Indeed, a different proof identifies
explicitly the nonclassical terms of the Heisenberg equations of
motion; these are operators which have no counterpart in the classical 
equations. Such terms (see below) yield a null expectation
value over states which are confined within an $n$-manifold, and also
negligible intermanifold contributions to the equations of motion (see main text).

Therefore, we consider next the case $\imath=2$, $\jmath=1$: 
\begin{equation}
\label{q_8}
	\begin{split}
	{\hat y}{\hat p}_{x} 
	& 
	= -i {\hat y} \left[ {\hat p}_{y}, {\hat L}{_z} \right] \\
	& 
	= -i {\hat y} {\hat p}_{y} {\hat L}_{z} 
	+ i \left\{ \left[ {\hat y} , {\hat L}_{z} \right] {\hat p}_{y} 
	+ {\hat L}_{z} {\hat y} {\hat p}_{y} \right\} \\
	&
	= - {\hat x} {\hat p}_{y} + i \left\{ {\hat L}_{z} {\hat y} {\hat p}_{y} 
	- {\hat y} {\hat p}_{y} {\hat L}_{z} \right\} \; .
	\end{split}
\end{equation}
We must then show that the expectation value over $| \psi_{n} \rangle$
of the operator within curly brackets 
vanishes, that is:
\begin{equation}
\label{q_9}
	\langle \psi_{n} |
	{\hat L}_{z} {\hat y} {\hat p}_{y} 
	- {\hat y} {\hat p}_{y} {\hat L}_{z} 
	| \psi_{n} \rangle = 0 \; .
\end{equation}
Clearly, the state $| \psi_{n} \rangle$ can be written as:
\begin{equation}
\label{q_12}
	| \psi_{n} \rangle =
	\sum_{ \ell , m } C_{n} ( \ell , m ) | n \ell m \rangle  \; ,
\end{equation}
where the $C_{n}( \ell , m)$'s are some general coefficients, possibly
complex. By substituting the expansion of Eq. (\ref{q_12}) in 
the expectation value of Eq. (\ref{q_9}) one has:
\begin{equation}
\label{q_12a}
	\begin{split}
	&
	\langle \psi_{n} |
	{\hat L}_{z} {\hat y} {\hat p}_{y} 
	- {\hat y} {\hat p}_{y} {\hat L}_{z} 
	| \psi_{n} \rangle \\
	&
	=
	\sum_{\ell^{\prime} , m^{\prime}} \! \sum_{\ell , m} 
	{\bar C}_{n} ( \ell^{\prime} , m^{\prime} )
	C_{n} ( \ell , m ) 
	\Big{\{}
	m^{\prime} \langle n \ell^{\prime} m^{\prime} | {\hat y} {\hat p}_{y} 
	| n \ell m \rangle \\
	&
	-
	m \langle n \ell^{\prime} m^{\prime} | {\hat y} {\hat p}_{y} 
	| n \ell m \rangle 
	\Big{\}} \; ,
	\end{split}
\end{equation}
where ${\bar C}_{n} ( \ell^{\prime} , m^{\prime} )$ denotes the 
complex conjugate.
On the other hand, from Eq. (\ref{q_7}) it follows that:
\begin{equation}
\label{q_10}
	\begin{split}
	\langle n \ell^{\prime} m^{\prime} |
	{\hat y} {\hat p}_{y} 
	&
	+ {\hat p}_{y} {\hat y} | n \ell m \rangle \\
	&
	= \langle n \ell^{\prime} m^{\prime} |
	2 {\hat y} {\hat p}_{y} + \left[ {\hat p}_{y} , {\hat y} \right]
	| n \ell m \rangle = 0 \; ,
	\end{split}
\end{equation}
and therefore the matrix elements of ${\hat y} {\hat p}_{y}$ are:
\begin{equation}
\label{q_11}
	\langle n \ell^{\prime} m^{\prime} |
	{\hat y} {\hat p}_{y} 
	| n \ell m \rangle
	= \frac{i}{2} \delta_{\ell^{\prime} \ell } \delta_{ m^{\prime} m } \; .
\end{equation}
By inserting the matrix elements 
of Eq. (\ref{q_11}) in the double sum of Eq. (\ref{q_12a})
it is easy to see that each term within 
curly brackets vanishes exactly, and therefore
the identity of Eq. (\ref{q_9}) is proved.

Next, we consider the case $\imath=3$, $\jmath=1$. One has:
\begin{equation}
\label{q_13}
	\begin{split}
	{\hat z}{\hat p}_{x} 
	& 
	= i {\hat z} \left[ {\hat p}_{z}, {\hat L}{_y} \right] \\
	& 
	= i {\hat z} {\hat p}_{z} {\hat L}_{y} 
	- i \left\{ \left[ {\hat z} , {\hat L}_{y} \right] {\hat p}_{z} 
	+ {\hat L}_{y} {\hat z} {\hat p}_{z} \right\} \\
	&
	= - {\hat x} {\hat p}_{z} - i \left\{ {\hat L}_{y} {\hat z} {\hat p}_{z} 
	- {\hat z} {\hat p}_{z} {\hat L}_{y} \right\} \; ,
	\end{split}
\end{equation}
and so we must prove that:
\begin{equation}
\label{q_14}
	\langle \psi_{n} |
	{\hat L}_{y} {\hat z} {\hat p}_{z} 
	- {\hat z} {\hat p}_{z} {\hat L}_{y} 
	| \psi_{n} \rangle = 0 \; .
\end{equation}
We use:
\begin{equation}
\label{q_15}
	{\hat L}_{y} = \frac{1}{2 i } \left( 
	{\hat L}_{+} - {\hat L}_{-} \right) \; ,
\end{equation}
and also:
\begin{eqnarray}
\label{q_16}
	{\hat L}_{+} | n \ell m \rangle & = & 
	\sqrt{ ( \ell - m ) ( \ell + m + 1 ) } | n \ell m + 1 \rangle \nonumber \\
	{\hat L}_{-} | n \ell m \rangle & = & 
	\sqrt{ ( \ell + m ) ( \ell - m + 1 ) } | n \ell m - 1 \rangle  \; ,
\end{eqnarray}
and the expectation value of Eq (\ref{q_14}) becomes:
\begin{equation}
\label{q_17}
	\begin{split}
	&
	\langle \psi_{n} |
	{\hat L}_{y} {\hat z} {\hat p}_{z} 
	- {\hat z} {\hat p}_{z} {\hat L}_{y} 
	| \psi_{n} \rangle \\
	& =
	\frac{1}{ 2 i }
	\sum_{ \ell^{\prime} , m^{\prime} } 
	\sum_{ \ell , m }
	{\bar C}_{n} ( \ell^{\prime} , m^{\prime} )
	C_{n} ( \ell , m ) \\
	&
	\times 
	\left\{
		\sqrt{ ( \ell^{\prime} + m^{\prime} ) ( \ell^{\prime} - m^{\prime} + 1 ) }
		\langle n \ell^{\prime} m^{\prime} - 1 |
		{\hat z} {\hat p}_{z} 
		| n \ell m \rangle  
	\right. \\
	&
	- \sqrt{ ( \ell^{\prime} - m^{\prime} ) ( \ell^{\prime} + m^{\prime} + 1 ) }
	\langle n \ell^{\prime} m^{\prime} + 1 |
	{\hat z} {\hat p}_{z} 
	| n \ell m \rangle  \\
	&
	- \sqrt{ ( \ell - m ) ( \ell + m + 1 ) }
	\langle n \ell^{\prime} m^{\prime} |
	{\hat z} {\hat p}_{z} 
	| n \ell m + 1\rangle \\
	&
	\left.
		+ \sqrt{ ( \ell + m ) ( \ell - m + 1 ) }
		\langle n \ell^{\prime} m^{\prime} |
		{\hat z} {\hat p}_{z} 
		| n \ell m - 1\rangle 
	\right\} \; .
	\end{split}
\end{equation}
Clearly, the matrix elements of ${\hat z}{\hat p}_{z}$ are also
given by Eq. (\ref{q_11}), and by inserting that result in Eq. (\ref{q_17})
it is easy to verify that once again the expression within 
curly brackets vanishes.

Finally, an essentially similar argument proves that the identity of Eq. 
(\ref{q_6}) holds also for ${\hat z}{\hat p}_{y}$, which completes
our proof.
%
%%%%%%%%%%%%%%%%%%%%%%%%%%%%%%%%%%%%%%%%%%%%%%%%%%%%%%%%%%%%%%%%%%%%%%%%%%%%%%%%%%%%%%%%%%%%%%%

\section{Time averaging of the intermanifold dynamics}

In this appendix we evaluate explicitly the time averages over a Kepler period
$T_{K}$ 
of the intermanifold contributions to the equations of motion for the
quantum expectation values.

We begin with:
\begin{equation}
\label{q_70}
	\left\langle
	{e} 
	^{ \mp i ( E_{\imath} - E_{\jmath} ) t } 
	\right\rangle _{K} =
	\frac{1}{T_{K}}
	\int_{0}^{T_{K}} 
	\!\!
	{e} ^{ \mp i ( E_{\imath} - E_{\jmath} ) t } dt 
\end{equation}
The integral of Eq. (\ref{q_70}) is easily evaluated, and one has:
\begin{equation}
\label{q_71}
	\frac{1}{T_{K}}
	\int_{0}^{T_{K}} 
	\!\!
	{e} ^{ \mp i ( E_{\imath} - E_{\jmath} ) t } dt 
	= \pm i
	\frac{ {e} ^{ \mp i ( E_{\imath} - E_{\jmath} ) T_{K} }  - 1 }
	{  ( E_{\imath} - E_{\jmath} ) T_{K}  }
\end{equation}
However, the energy difference $E_{\imath} - E_{\jmath}$ is:
\begin{equation}
\label{q_72}
	\begin{split}
	E_{\imath} - E_{\jmath}
	& =
	- \left(
	\frac{1}{ 2 \imath^{2} } - \frac{1}{ 2 \jmath^{2} }
	\right) \\
	& =
	\frac{ \Delta_{\imath,\jmath} }{ \jmath^{3} } -
	\frac{ 3 \Delta_{\imath,\jmath} ^{2} }{ 2 j } \frac{1}{ \jmath^{3} }
	+ O \left( \frac{ \Delta_{\imath,\jmath} ^{3} }{ \jmath^{5} } \right) \; ,
	\end{split}
\end{equation}
where $\Delta_{\imath,\jmath} = \imath - \jmath$.
We then use:
\begin{equation}
\label{q_72a}
	\jmath = {\bar n} \left( 1 + \frac{ \Delta_{\jmath,{\bar n}} }{ {\bar n} } \right) \; ,
\end{equation}
where $\Delta_{\jmath,{\bar n}} = \jmath - {\bar n}$ 
and ${\bar n}$ is the principal quantum number 
of the hydrogenic manifold which carries the largest weight in the state.
The energy difference between two manifolds can then be rewritten as:
\begin{equation}
\label{q_72b}
	E_{\imath} - E_{\jmath} =
	\frac{ \Delta_{\imath,\jmath} }{ {\bar n}^{3} }
	\left\{ 1 - \frac{3}{{\bar n}}
	\left( \Delta_{\jmath,{\bar n}} - \frac{1}{2} \Delta_{\imath,\jmath} \right)
	\right\}
	+ O \left( \frac{ \Delta^{3} }{ {\bar n}^{5} } \right) \; .
\end{equation}
where $\Delta^{3}$ (i.e., with no indexes) stands for the product of any three
$\Delta$'s regardless of the indices. 
The Kepler period is:
\begin{equation}
\label{q_73}
	T_{K} = 2 \pi {\bar n}^{3}
\end{equation}
Substituting the results of Eq. (\ref{q_72b}) and Eq. (\ref{q_73}) in
Eq. (\ref{q_71})
one obtains:
\begin{equation}
\label{q_75}
	\begin{split}
	&
	\frac{1}{T_{K}}
	\int_{0}^{T_{K}} 
	\!\!
	{e} ^{ \mp i ( E_{\imath} - E_{\jmath} ) t } dt 
	\\
	& =
	\frac{3}{{\bar n}} \left( \Delta_{{\bar n},\jmath} - \frac{1}{2} 
	\Delta_{\imath,\jmath} \right) +
	O \left( \frac{ \Delta^{2} }{ {\bar n}^{2} } \right) \; .
	\end{split}
\end{equation}
Note that to the leading order in $\Delta / {\bar n} $ the result does not
depend on the sign of the exponent; in fact, the leading term of the
right hand side of Eq. (\ref{q_75}) can 
be cast in a more symmetric form:
\begin{equation}
\label{q_76}
	\Delta_{{\bar n},\jmath} - \frac{1}{2} \Delta_{\imath,\jmath}  =
	\Delta_{{\bar n},\imath} - \frac{1}{2} \Delta_{\jmath,\imath}  =
	{\bar n} - \frac{1}{2} \left( \imath + \jmath \right)
\end{equation}
which concludes the calculation of the first time average.

Incidentally, by inverting to the leading order the 
expression of Eq. (\ref{q_72b}) we obtain a result which 
we used in the main text of this paper:
\begin{equation}
\label{q_75a}
	\frac{ 1 }{ E_{\imath} - E_{\jmath} } =
	\frac{ {\bar n}^{3} }{ \Delta_{\imath,\jmath} }
	\left\{
	1 + \frac{3}{{\bar n}}
	\left( \Delta_{\jmath,{\bar n}} - \frac{1}{2} \Delta_{\imath,\jmath} \right)
	\right\}
	+ O \left( \frac{ \Delta^{3} }{ {\bar n}^{5} } \right) \; . 
\end{equation}

Next we evaluate:
\begin{equation}
\label{q_77}
	\left\langle
	i t \ {e} ^{ - i ( E_{\imath} - E_{\jmath} ) t }
	\right\rangle _{K} =
	\frac{ i }{T_{K}}
	\int_{0}^{T_{K}} 
	\!\!
	t \ {e} ^{ - i ( E_{\imath} - E_{\jmath} ) t } dt 
\end{equation}
Once again, the integral is straightforward:
\begin{equation}
\label{q_78}
	\begin{split}
	&
	\frac{ i }{T_{K}}
	\int_{0}^{T_{K}} 
	\!\!
	t \ {e} ^{ - i ( E_{\imath} - E_{\jmath} ) t } dt = 
	\frac{1}{ ( E_{\imath} - E_{\jmath} ) T_{K} } \\
	& \times
	\left\{
	i
	\frac{ {e} ^{ -  i ( E_{\imath} - E_{\jmath} ) T_{K} }  - 1 }
	{  ( E_{\imath} - E_{\jmath} ) }
	-
	T_{K} {e} ^{ - i ( E_{\imath} - E_{\jmath} ) T_{K} } \right\} \; .
	\end{split}
\end{equation}
Finally, inserting in Eq. (\ref{q_78})
the results of Eqs. (\ref{q_72}-\ref{q_73}) one obtains:
\begin{equation}
\label{q_79}
	\begin{split}
	&
	\frac{ i }{T_{K}}
	\int_{0}^{T_{K}} 
	\!\!
	t \ {e} ^{ - i ( E_{\imath} - E_{\jmath} ) t } dt 
	= 
	- \frac{ {\bar n}^{3} }{ \Delta_{\imath,\jmath} }
	{\bigg{\{} }
	1 \\
	& {\hspace{-0.8cm}} - 
	\frac{6}{{\bar n}} 
	\left( \Delta_{{\bar n},\jmath} - \frac{1}{2} \Delta_{\imath,\jmath} \right)
	\left( 1 + i \pi \Delta_{\imath,\jmath} \right) 
	+ O \left( \frac{ \Delta^{3} }{ {\bar n}^{2} } \right)  \bigg{\}} \ ,
	\end{split}
\end{equation}
which concludes our analysis.

\bibliography{strings,myself,chaology,rydberg,turgay,books,zeke,triangles,%
packets}
%
%%%%%%%%%%%%%%%%%%%%%%%%%%%%%%%%%%%%%%%%%%%%%%%%%%%%%%%%%%%%%%%%%%%%%%%%%%%%%%%

%%%%%%%%%%%%%%%%%%%%%%%%%%%%%%%%%%%%%%%%%%%%%%%%%%%%%%%%%%%%%%%%%%%%%%%%%%%%%%%
%
\begin{comment}
%

\clearpage

{\bf \centerline{Figure Captions}}

\vspace{1.0cm}
\noindent Figure {\ref{q_fig_1}}

\noindent%
Scaling of the intermanifold contribution with the principal
quantum number. 
In Fig (a) we plot the natural logarithm of the maximum 
magnitude of the intermanifold terms vs. the natural logarithm
of the principal quantum number, and the approximate straight 
line indicates a simple power-law scaling.
In Fig (b) we plot the slope of the line in Fig. (a), i.e.,
the exponent of the power-law, which is clearly converging to 
$\xi = 3$, thereby proving that in the semiclassical limit all
intermanifold contributions become negligible.

\clearpage

%
%%%%%%%%%%%%%%%%%%%%%%%%%%%%%%%%%%%%%%%%%%%%%%%%%%%%%%%%%%%%%%%%%%%%%%%%%%%%%%%
%
% Importing a Figure
%
\begin{figure}
\centerline{\psfig{file=q_fig_1.eps,height=12.0cm,angle=0}}
\caption{
}
\label{q_fig_1}
\end{figure}
%
%%%%%%%%%%%%%%%%%%%%%%%%%%%%%%%%%%%%%%%%%%%%%%%%%%%%%%%%%%%%%%%%%%%%%%%%%%%%%%%
%

\end{comment}

%%%%%%%%%%%%%%%%%%%%%%%%%%%%%%%%%%%%%%%%%%%%%%%%%%%%%%%%%%%%%%%%%%%%%%%%%%%%%%%

\end{document}